\documentclass[aps,pre,showpacs,showkeys]{revtex4-1}

\usepackage{geometry,amsmath,color,ulem, rotating}
\usepackage{amssymb}
\usepackage{graphicx,booktabs,array,fancyhdr,subfig,hyperref,sidecap} 
\usepackage{tikz, appendix,caption}
\usetikzlibrary{shapes,arrows}
\begin{document}

\title{Swallowing a cellular automaton pill: predicting drug release from a matrix tablet}
\author{Ezra Buchla}
\email[]{e@catfact.net}
\affiliation{1515 Bates Place, Claremont, CA  91711, USA}
\author{Peter Hinow}
\email{hinow@uwm.edu}
\affiliation{Department of Mathematical Sciences, University of Wisconsin -- Milwaukee, P.O.~Box 413, Milwaukee, WI 53201, USA}
\author{Aisha N\'ajera}
\email{aisha.najera@cgu.edu}
\affiliation{Department of Mathematics, Claremont Graduate University, 710 N.~College Ave., Claremont, CA  91711, USA}
\author{ Ami Radunskaya}
\email{aer04747@pomona.edu}
\affiliation{Department of Mathematics, Pomona College, 610 N. College Ave., Claremont, CA  91711, USA}

\date{\today} 
\begin{abstract} Matrix tablets are drug delivery devices designed to release a drug in a controlled manner over an extended period of time.
We develop a cellular automaton (CA) model for the dissolution and release of a
water-soluble drug and excipient from a matrix tablet of water-insoluble polymer. Cells of the CA are occupied by
drug, excipient, water or polymer and the CA updating rules  simulate the dissolution of  drug and  excipient and the subsequent
diffusion of the dissolved substances. In addition we simulate the possible fracture of brittle drug and excipient powders
during the tablet compression and the melting of the polymer during a possible thermal curing process.  
Different stirring mechanisms that facilitate the transport of dissolved drug in the fluid in which the tablet is immersed are modeled in the water cells adjacent to the 
boundary of the tablet. We find that
our simulations can reproduce experimental drug release profiles. Our simulation tool can be used to streamline the 
formulation and production of sustained release tablets.
\end{abstract}
\maketitle

\section{Introduction}
The design of controlled release systems has been an active area of research in pharmaceutical science and industry for decades. The most common 
systems include coated systems, matrix tablets, eroding tablets and oral osmotic therapeutic systems \cite{Fredenberg,Zuleger,Casault_PhysA}. The present 
work focuses on the development of a mathematical model and simulation tool that describes the sustained drug release observed in matrix tablets. 
Matrix tablets are devices that deliver a drug in a controlled manner over an extended period of time. A preferred manufacturing method is to 
mix the drug with pharmaceutically inactive excipient and polymer powders.  This powder mixture is then 
compressed in a die at high pressures  (e.g.~$70-200\,MPa$) and may be cured for 8-24 hours at $40-70\,{}^\circ C$. Until now, pharmaceutical scientists were restricted to fabrication of experimental tablets to understand the influence of parameters such as the powder composition, 
the compaction pressure and the curing temperature and duration. As this is expensive and time 
consuming, the need for mathematical modeling and simulation tools becomes evident, see \cite{Lemaire,Villalobos_PhysA,Siepmann,DCDSB} for earlier works. 
The aim of this work is to provide a tool  that allows pharmaceutical scientists to mimic the release processes in matrix tablets.

In our earlier work \cite{DCDSB} we proposed two mathematical models for the release of a water-soluble drug from a 
polymer/excipient matrix tablet. The first model used a biased random walk on the contact graph of a random sphere packing.  
The second model was based on a  system of reaction-diffusion partial differential equations for the concentrations
of dissolved and undissolved drug and excipient, respectively. While the first discrete model predicted partial release of drug 
from the tablet in agreement with experimental observations \cite{DCDSB}, the second continuous model proved 
better at capturing a change from convex to concave  in several experimental release profiles. 
With the cellular automaton (CA) model presented here, we explicitly model the initial wetting process of the tablet 
when water passes through pores into the interior of the tablet, followed by the dissolution of the drug and the 
excipient. We also incorporate the fracture of the brittle drug and excipient powders during the compaction process and the 
partial melting of the polymer during an optional thermal treatment of the compressed tablet. This results in the formation of larger connected
regions of polymer that may entrap drug and prevent it from leaving the tablet. Finally, we also allow to simulate different stirring protocols that
aide the transport of dissolved and released drug away from the tablet \cite{Viegas}. We find that our model has the capacity to reproduce quantitatively experimental release curves of tablets formulated from different powder mixtures and subjected to different heating protocols. 

Cellular automata have shown their usefulness in many sciences since their introduction by Ulam and von Neumann about 70 years ago. Let us mention here only some applications to traffic networks \cite{Nowak_PRE85}, neural networks \cite{Goltsev_PRE81}, tumor growth \cite{Kavousanakis_PRE85} and statistical mechanics \cite{Wolfram1983}. They allow for greater detailed modeling of internal processes in ``cells'' that is generally not possible in coarse-grained partial differential equation models. The recent advances  in computational technology have made earlier restrictions to small numbers of cells less stringent, if not obsolete. 

The remainder of this paper is organized as follows. We describe in Section \ref{CAmod} the cellular automaton model in detail, paying special attention to the initialization process. Section \ref{results} contains several simulation results, where we explore the influences of the respective parameters. In Section we discuss our results and indicate future research. Appendix \ref{parameters} lists parameter values used in the simulations and instructions how to download and use our simulation tool.

\section{The cellular automaton  model}\label{CAmod}
\subsection{General description and program flow}

\begin{figure}[ht]
\includegraphics[width=90mm]{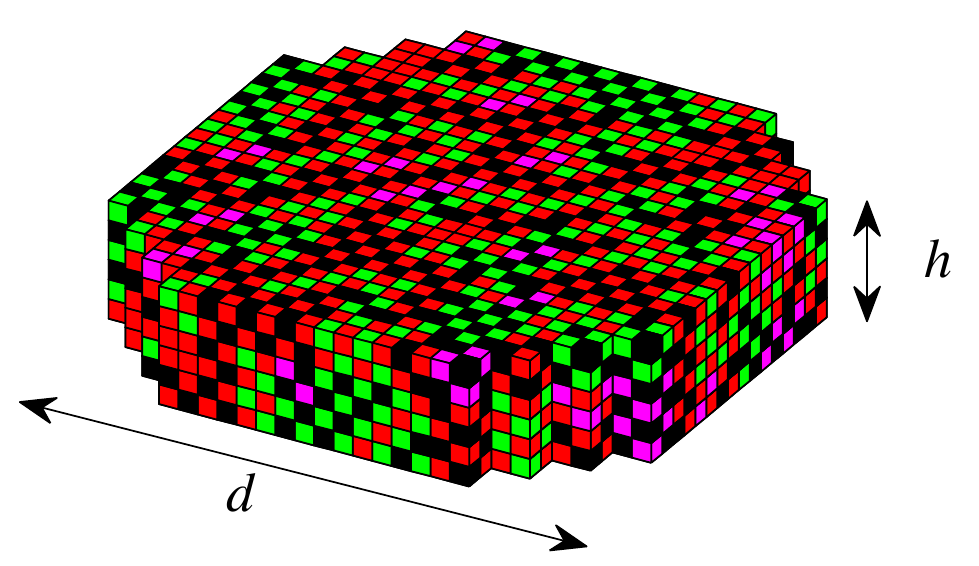} \hspace{5mm}  
\includegraphics[width=45mm]{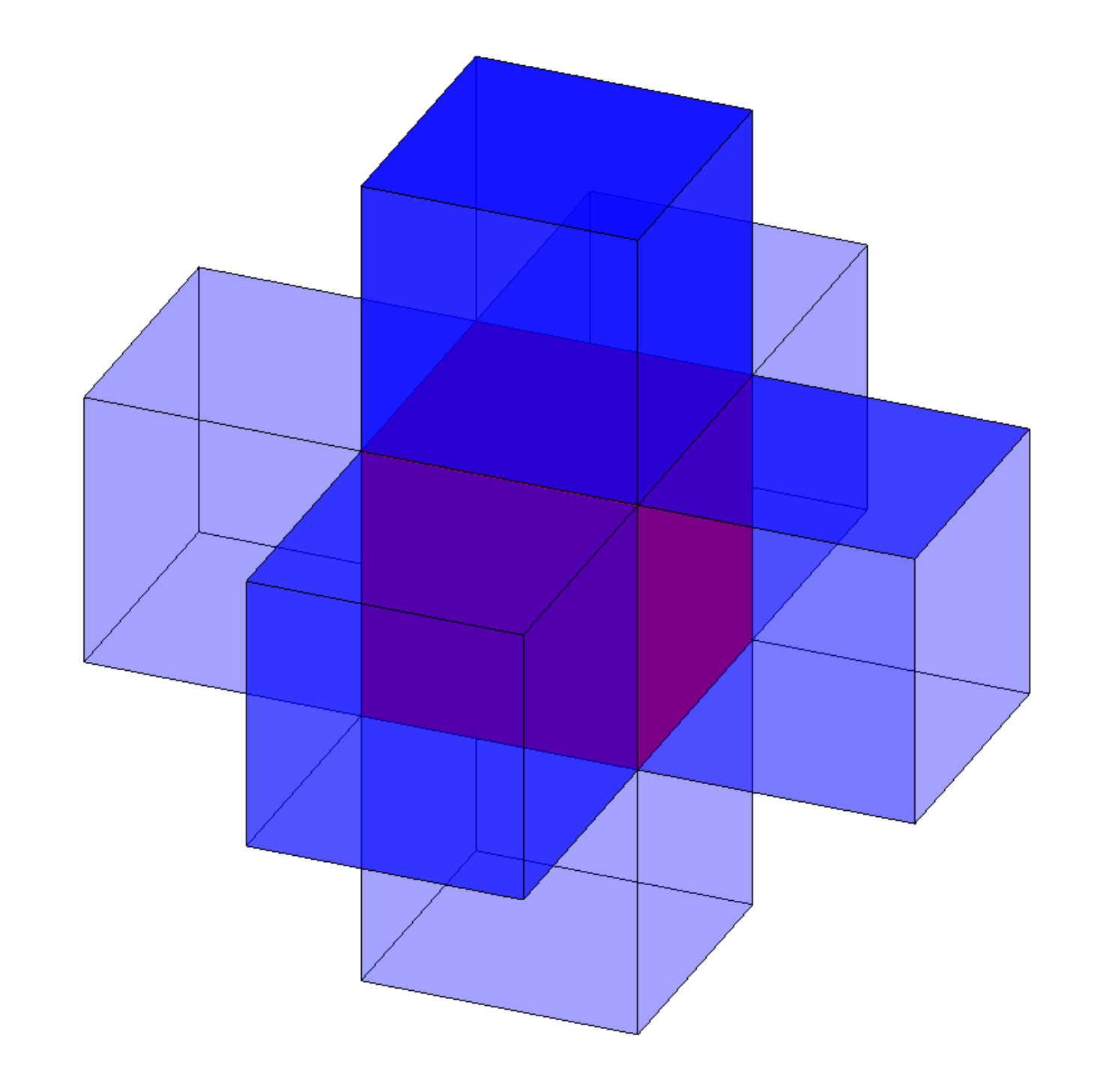}
\caption{ (Left) The tablet is represented by a cubic lattice in a cylindrical domain. Each grid element, indicated here by a colored cube, can be in one of five states:  drug (purple),  polymer (red), excipient (green), empty (black) and  water (blue). This tablet is surrounded by a layer of water cells (not shown here). (Right)  Each element interacts with its 6 closest von Neumann neighbors.}\label{TabletFig}
\end{figure}

The cellular automaton recreates the three-dimensional cylindrical geometry of the matrix tablet. 
Each particle in the tablet is represented by a vertex  in a cubic lattice that fills out a  cylindrical domain.  
The diameter  of the cylinder and height as a fraction of the diameter are specified by the user. A vertex interacts 
with its 6 closest von Neumann neighbors by a set of rules designed to imitate the actual physics of the release process,
see Figure \ref{TabletFig}. 
 
Every  cell  can be in one of five states, namely \textit{drug} (D), \textit{polymer} (P), \textit{excipient} (X), \textit{empty} (E) and \textit{water} (W).  In the initialization phase the cells of the lattice are randomly assigned a state so that the distribution of cells of type D, P and X are those prescribed by the composition of the powder mixture, with the possible formation of a polymer shell under thermal treatment. Empty cells are created during the heating and compression phase.  Cells of type W surround the tablet at all times. The wet cells are further characterized by two quantities, the concentration of drug and of excipient, respectively. These quantities are real numbers between 0 and 1, where a concentration of 1 corresponds to a saturated solution. Naturally, the concentrations change as water travels into the tablet and solid particles dissolve. To avoid clumsy triple index notation, we simply write $c_D(x,t)$ and $c_X(x,t)$  for the concentration of drug  and  excipient, respectively,  in the cell located at vertex $x$ of the lattice at time $t$. Throughout the paper, let $N(x)$ denote the von Neumann neighborhood of vertex $x$.

After initialization of the tablet according to the user's specification, the tablet's  state is repeatedly updated until either a maximum number of  iterations have been executed or the amount of released drug ceases to  change significantly.   Each iteration of the update rules  advances the simulation clock by a certain fixed amount of time which depends on the model parameters. Finally, we also model the transport of dissolved drug in the boundary layer surrounding the tablet. The details of the simulation are described in the following sections.

\begin{figure}[ht]
\includegraphics[width=125mm]{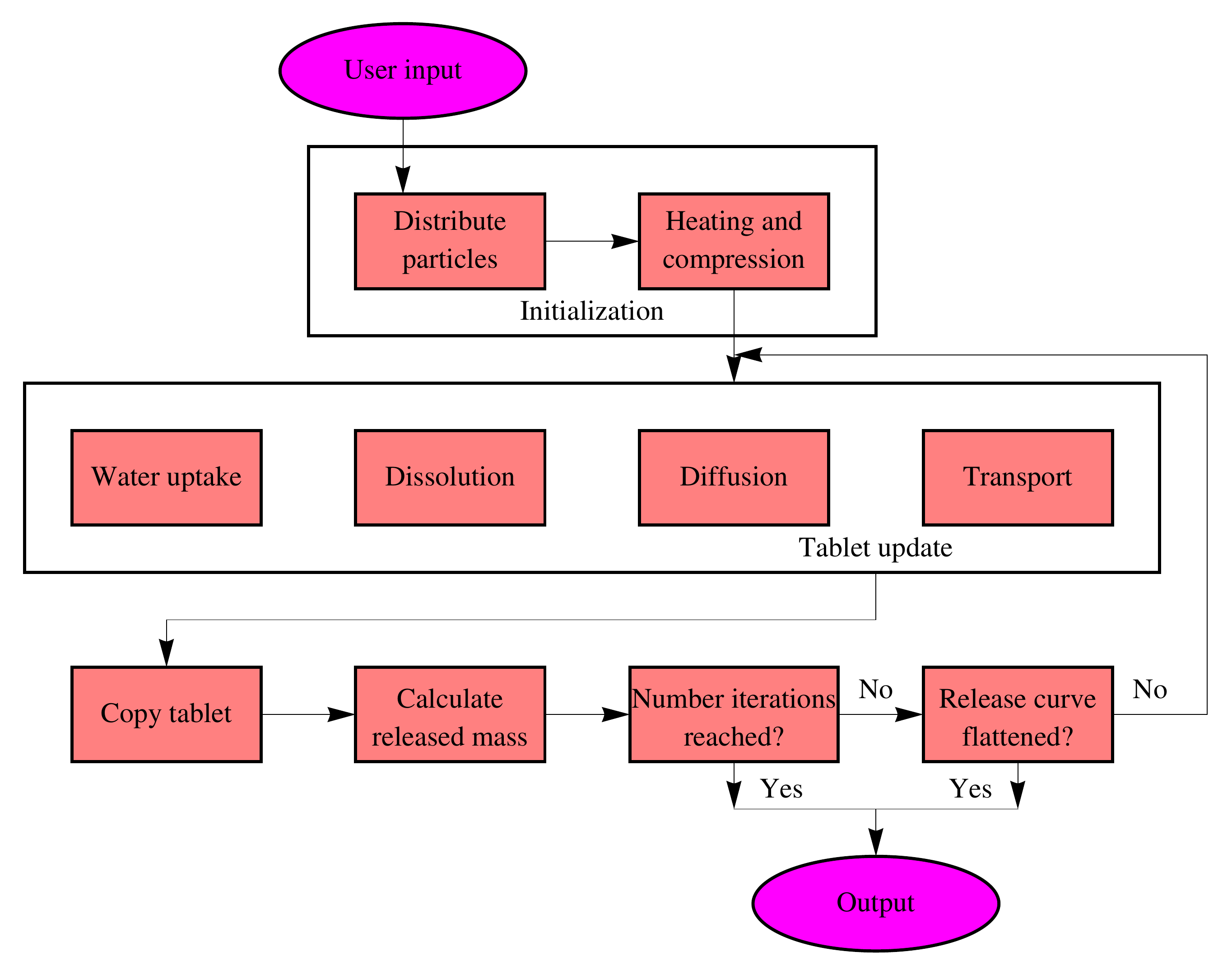}
\caption{Schematic representation of the flow. During each update of the tablet, a new tablet state is created from the existing one which is afterwards cleared to free memory.}\label{flow}
\end{figure}

\subsection{Initialization}

The composition of the tablet is determined by the fractions of drug, polymer and excipient in the powder, which we denote here by $P_D, P_P$ and $P_X$, respectively.  
Thermal treatments and compaction studies on Eudragit-RLPO matrix tablets  revealed that the particles of the powder mixture move and diffuse during the heating process \cite{Azarmi,Chatterjee2}, coalescing the polymer particles and forming matrices with decreased porosity and increased tortuosity. Porosity and tortuosity are important factors that affect the dissolution and release rate, as well as the possible formation of a thin polymer film on the surface of the tablet \cite{Azarmi}.
The release kinetics from laboratory studies show that, as the amount of polymer is increased, drug release from heated tablets approaches ``first-order kinetics'', i.e.~nearly linear release with respect to time, see \cite[Figure 7]{DCDSB}.

\begin{figure}
\hbox{\includegraphics[height = 1.84in]{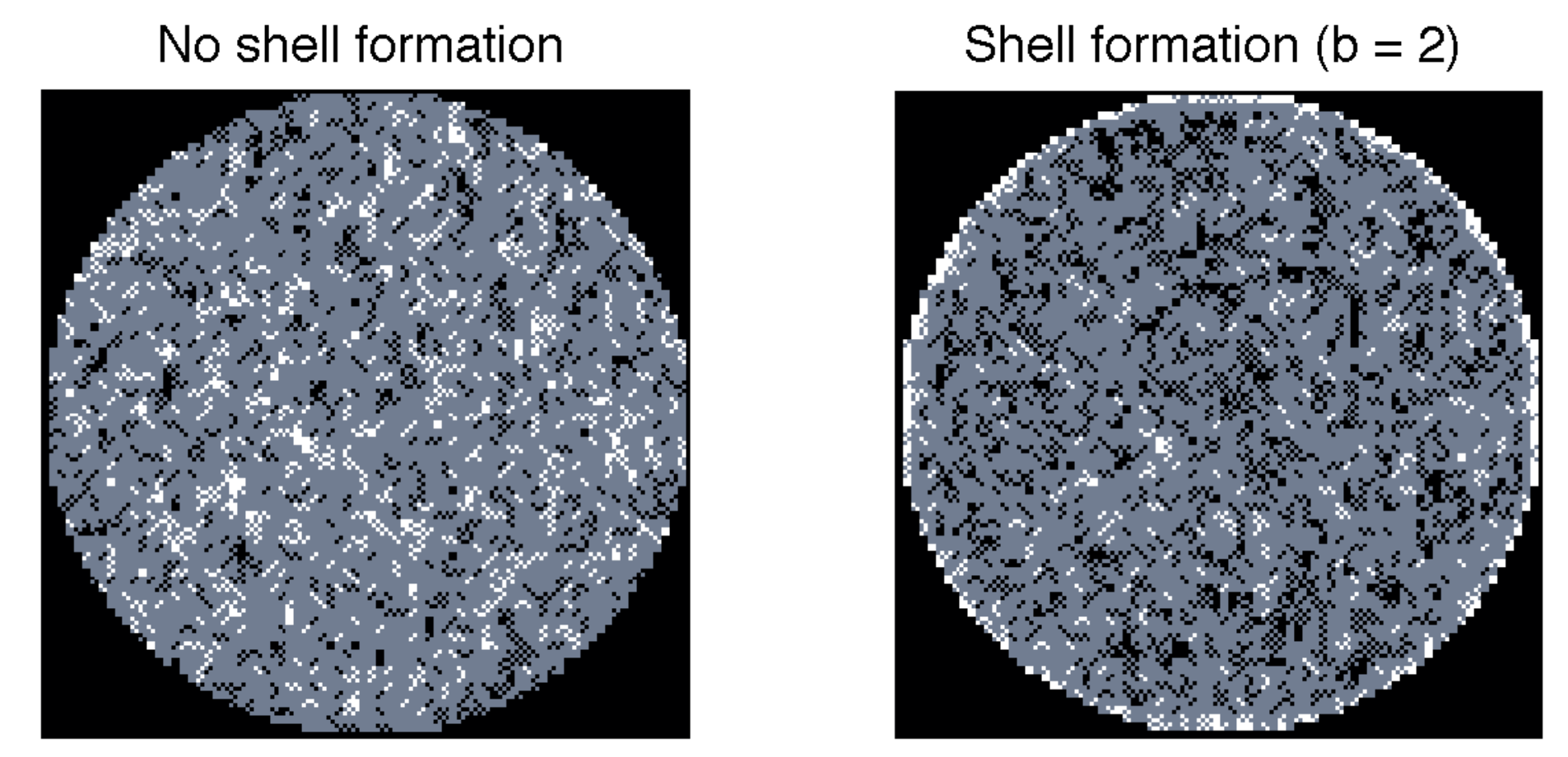} \quad \includegraphics[height = 1.78in]{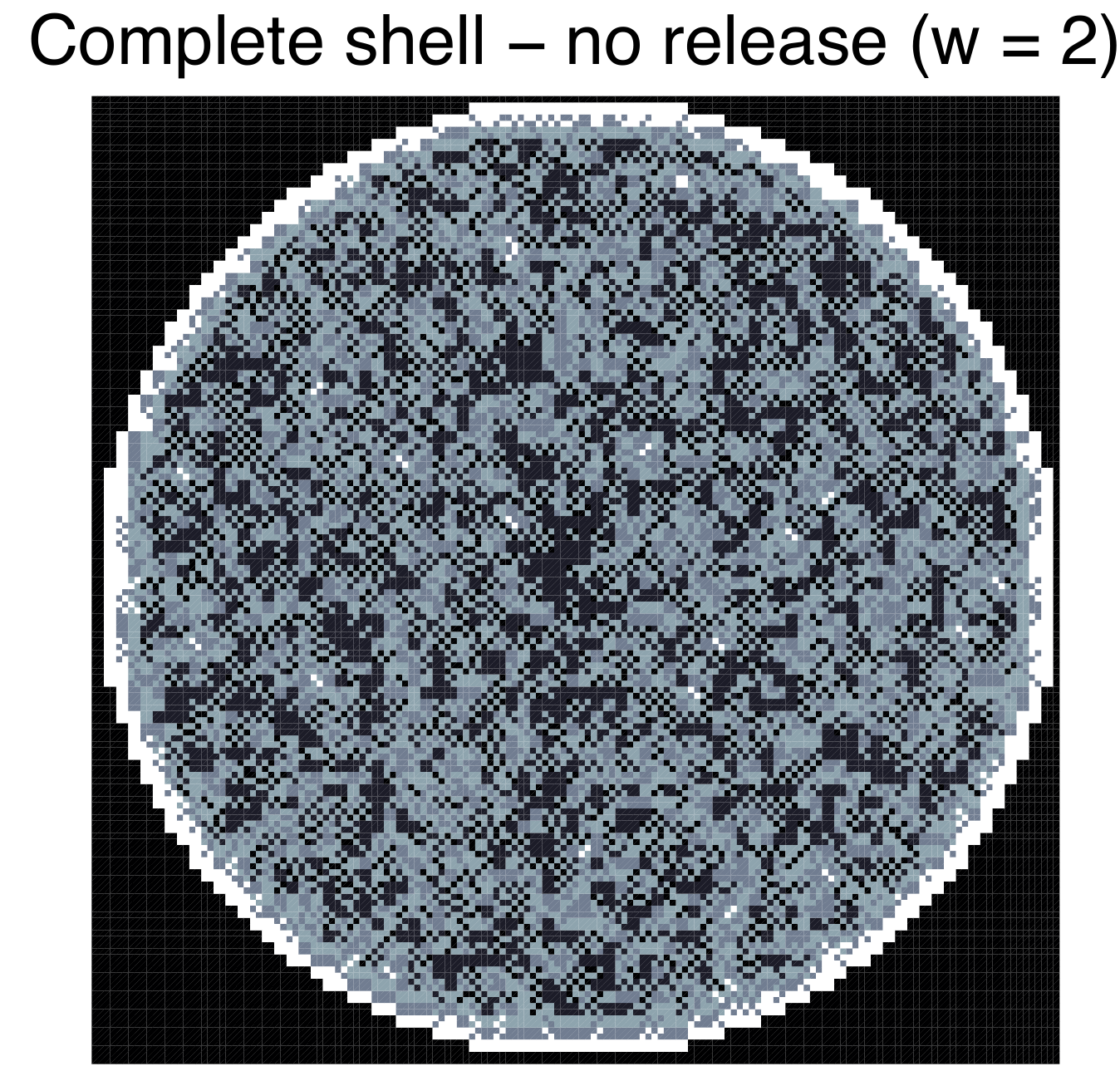}}
\caption{ An example of the formation of a polymer shell after thermal treatment.  The tablet contains 50\% polymer.   Polymer cells are shown in white, drug and excipient
cells are shown in black, and gray cells represent mixtures of polymer and
other particles.  (Left) $\mathtt{b} = 1, \mathtt{w = 1}$ (default) , Center:  $\mathtt{b} = 2, \mathtt{w} = 1$, giving a shell with some ``holes". (Right) $\mathtt{b} = 2, \mathtt{w} = 2$, giving a complete shell allowing no release. All other parameters are given in Table \ref{params}.} \label{Fig:shell}
\end{figure}

We simulate the creation of a film, or shell on the outside of the tablet as an effect of heating, as suggested in \cite{Azarmi}.  In the model, polymer particles can move towards the surface during thermal treatment, resulting in a higher chance of polymer being found on the tablet surface than in the interior.  The user specifies the thickness of the shell layer  ({\tt w} in the program) and an imbalance factor ${\tt b}\ge1$ that represents the preference of polymer particles to locate in the shell over the remainder of the tablet.  The distribution of particles is then computed as follows.  Let  $n_S$ denote the number of cells (of any type) in the shell layer, and let $n_T$ denote  the number of cells in the remainder of the tablet. To simulate shell formation, a number of $n_{PS} = n_S\min\{ b P_D ,1 \}$  polymer cells are randomly assigned to cells in the shell layer of the tablet.  The remaining  polymer cells, whose number is $n_{PT} = P_D   (n_S+ n_T)  - n_{PS}$  are randomly assigned to the interior of the tablet.  All unoccupied tablet cells (including cells in the shell layer) are then reshuffled, and are assigned to drug or excipient states, according to their prescribed fractions in the tablet. It should be emphasized that setting ${\tt b}=1$ results in no shell formation.  See Figure \ref{Fig:shell} for an illustration of simulated shell formations, and Figure \ref{Fig:change}, top left panel, for a comparison of release curves with and without shells.

\begin{figure}
\begin{tikzpicture}[node distance = 1.5in, auto]
    \node[shape = ellipse, draw, color = red]  (init) {Initialize};
    \node[right of = init, label = {above: Initial Drug Cell}] (initialcell) {\includegraphics[scale = .2]{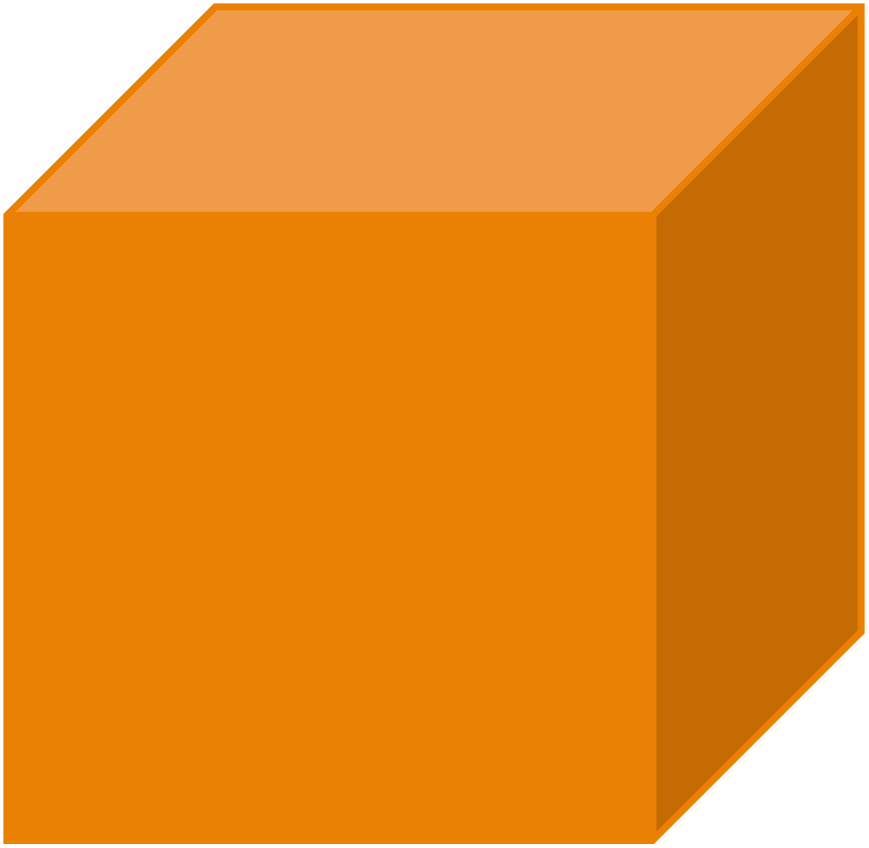}};
    \node[right of = initialcell, label = {above: Subdivided Cell with Empty Sub-cells}, node distance = 2.5in] (subcell) {\includegraphics[scale = .2]{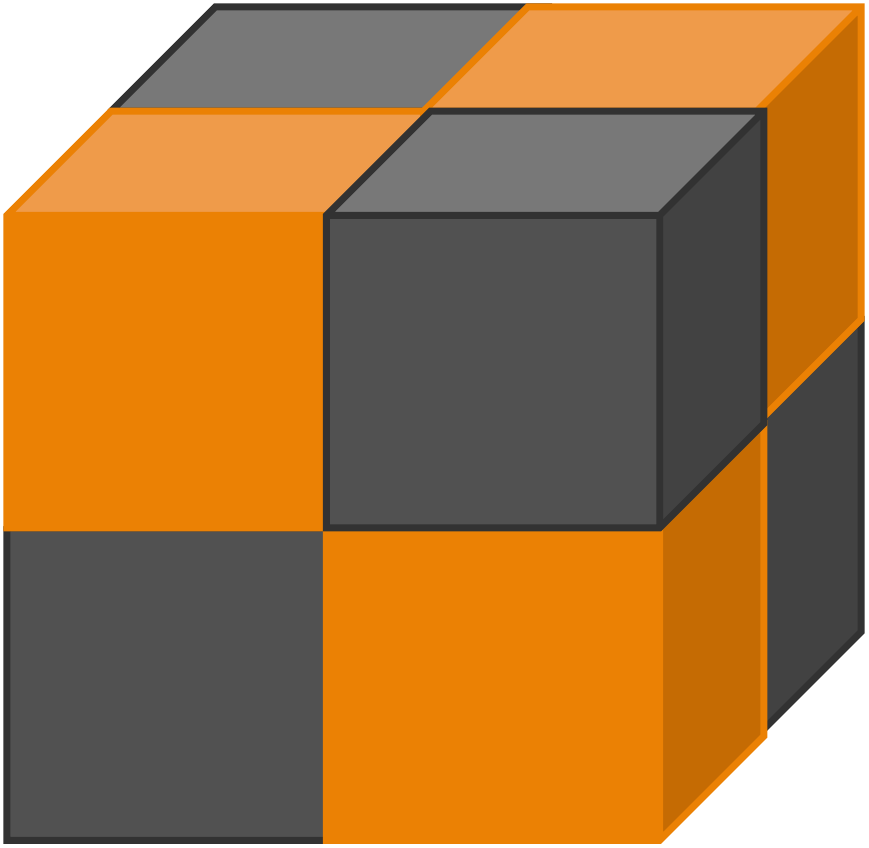}};
    \node[below of = init, label = {above:Polymer Before}]  (polybefore) {\includegraphics[scale = .2]{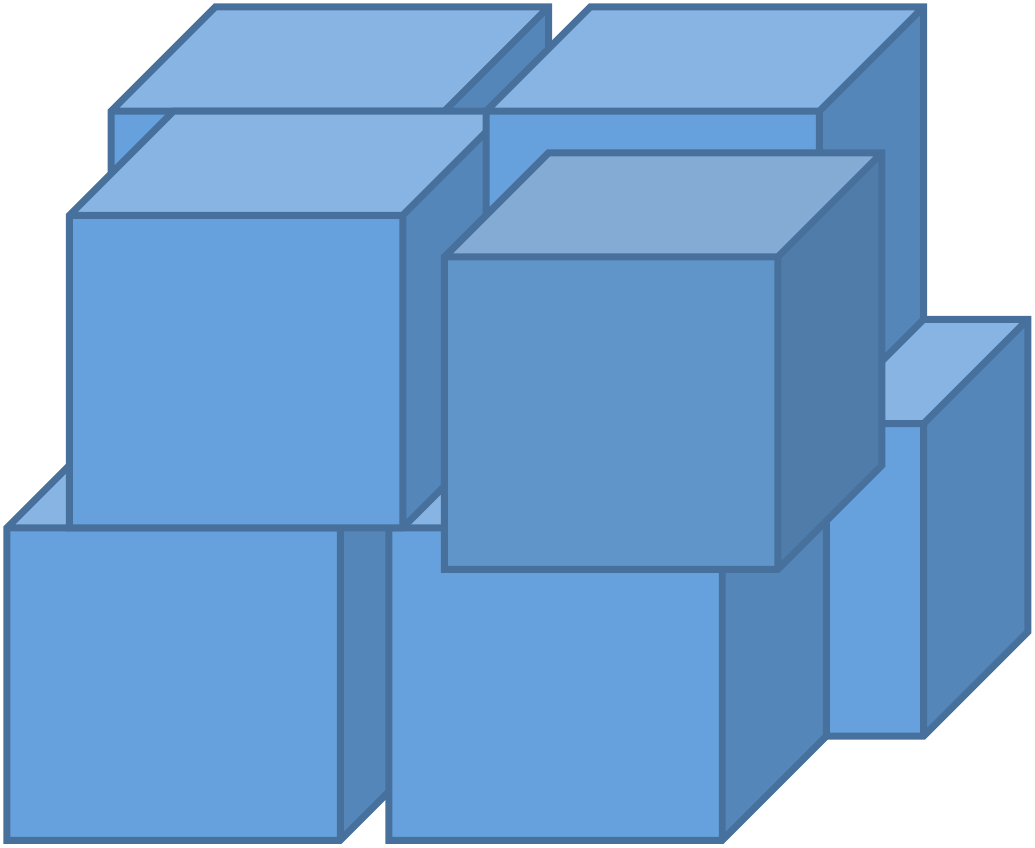}};
    \node [right of=polybefore, label = {above:Drug Before}] (drugbefore) {\includegraphics[scale = .2]{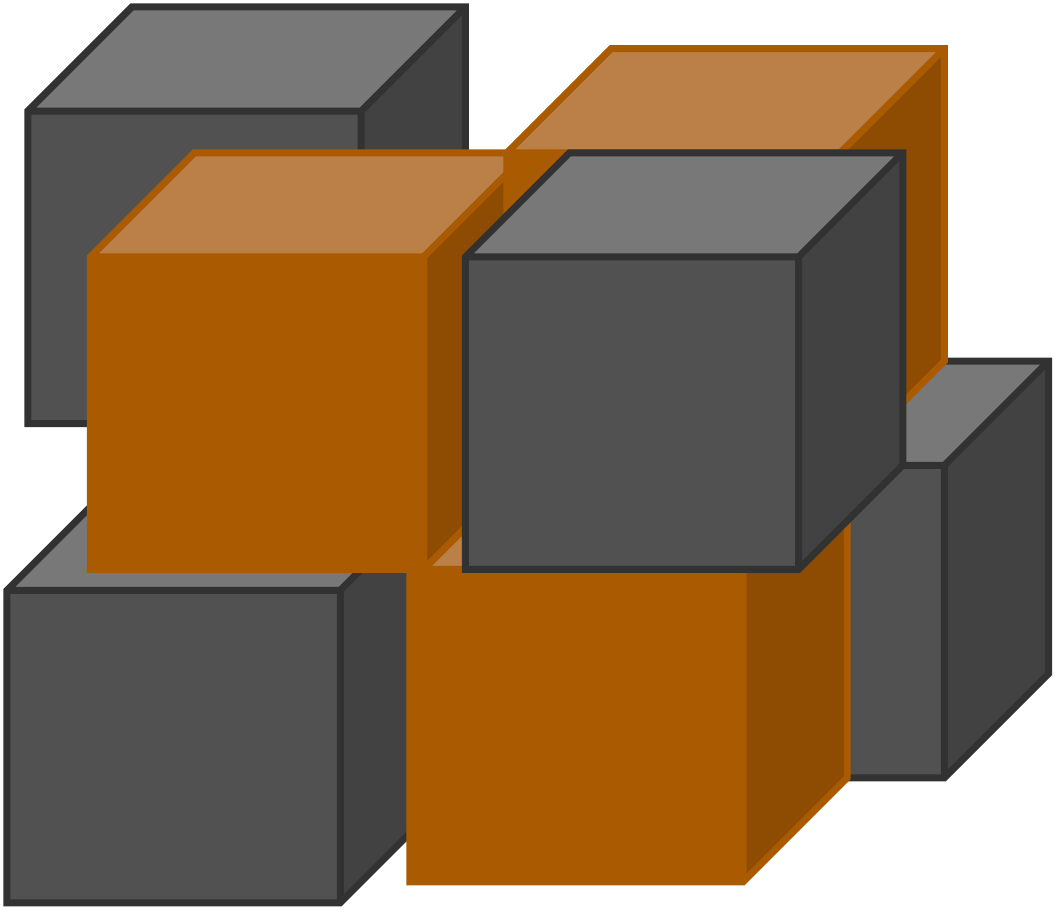}};
    \node [right of = drugbefore, node distance = 2.4in, label = {above: Polymer After}] (polyafter) {\includegraphics[scale = .2]{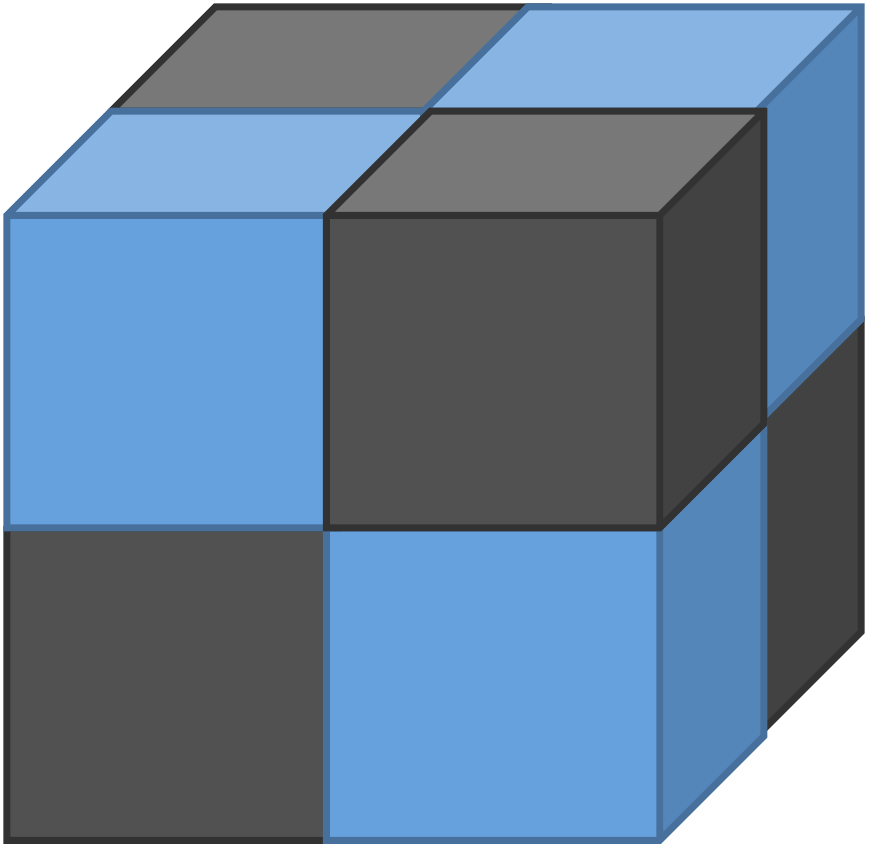}};
    \node [right of = polyafter, label = {above: Drug After}, node distance = 1.1in] (drugafter) {\includegraphics[scale = .2]{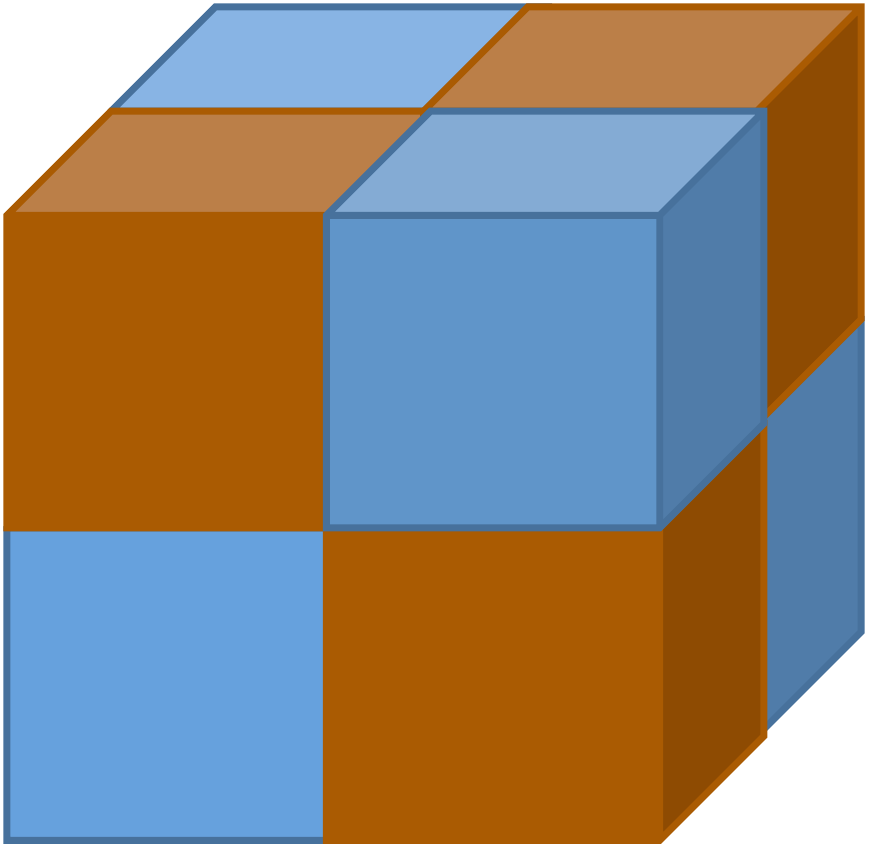}};
\path[->] (init) edge (initialcell);
\path[->, very thick, black] (initialcell) edge (subcell);
\path[->] (polybefore.center) edge[bend right] (drugbefore);
\path [->] (drugbefore) edge[bend right] (polybefore.east);
\path [->, very thick, double, red] (drugbefore) edge node {\it sub-cells swapped} (polyafter);
\end{tikzpicture}
\caption{Diagram representing the grid subdivision and ``swapping" algorithm used to simulate polymer deformation and spreading.  Each grid element is subdivided into 8 sub-cells, and empty (black) sub-cells are created in drug or excipient grid elements (brown).  Polymer sub-cells (blue) are swapped with empty (black) sub-cells during the simulation of thermal treatment.}
\label{Fig:compress}
\end{figure}
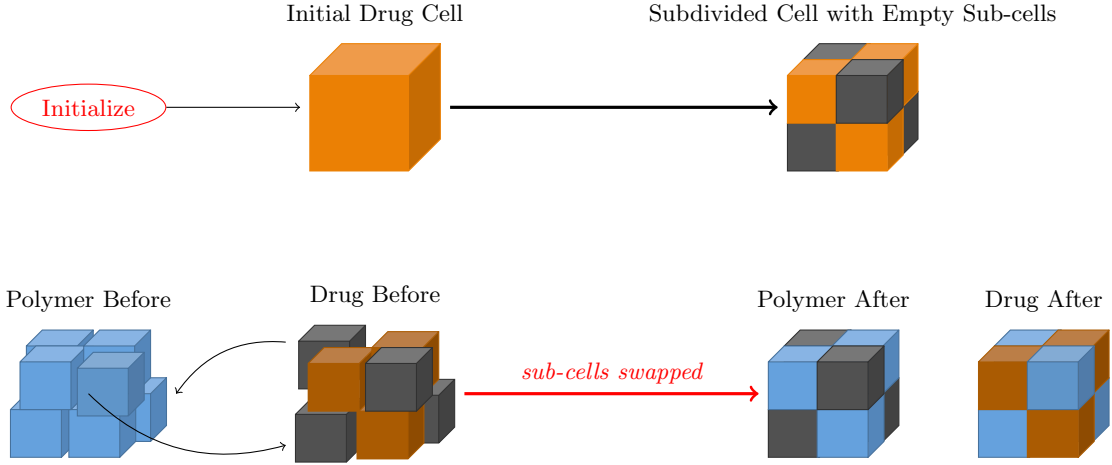

Experimental release curves at various polymer concentrations, curing temperatures and compression pressures show two distinctive characteristics that must be matched by a mathematical model. These are the initial slope of the release curve, and the amount of drug that is not released after a certain period of time, i.e.~the ``trapped" drug mass, when the release curve has flattened.  In order get the observed range of trapped drug mass,  the model must reflect the deformation of particles under pressure and the fusing of polymer particles under thermal treatment (similar to sintering).  We emulate these effects by refining the original grid and by allowing polymer particles to move into adjacent grid elements, increasing tortuosity and the likelihood of trapping.  First, each grid element is subdivided into 8 subcells of equal size.  If the grid element was in state $D$ or $X$, corresponding to drug or excipient particles, then the states of half of these subcells are changed to $E$, empty.  The new empty,  state $E$ subcells are located in opposing diagonal elements.  This subdivision of the cells corresponds to the breaking down of brittle particles under pressure.  Since polymer particles do not break under pressure, they are subjected to a different ``heating" rule.  Thermal treatment is modeled by swapping void subcells and polymer subcells if a polymer particle is located adjacent to a drug or excipient particle.  If a polymer particle is adjacent to several non-polymer particles, one of these neighbors is chosen randomly and subcells are swapped.  The processes of subdivision and subcell swapping are shown graphically in Figure \ref{Fig:compress}.   The result is a virtual tablet that resembles an actual tablet after thermal treatment.  
 
After initialization, the simulation domain is a cylindrical subregion of a cubic lattice, where each grid cell has side length equal to $L$. If compression is used, then the side length is half the size provided by the user $L = {\tt y}/2$.  The update rules are applied to this computational domain, so that a grid cell is one of these cells of length $L$, and the number of cells in one direction of the cubic lattice is ${\tt n}/L$. See Figure \ref{flow} for a schematic overview of the program flow.

\subsection{Updating rules}

The cell states are updated according to their present status and the following rules.   Table \ref{params} lists all user-defined parameters and their default values.
\begin{enumerate}
\item {\it Wetting.} For a cell of type E we simulate its filling with water from neighboring
cells of type W due to capillarity. If a cell of type E has at least one neighbor of type W, then its status is changed to W, otherwise no action is taken.

\item {\it Dissolution.}  A cell of type D or X that is in contact with cells of type W, can be replaced by a W cell. Assume that cell $x$ is in state  D,  the case that a cell is in state  X follows the obvious modifications. Denote by $n_W(x)$ the number of neighboring water cells of cell $x$ and by $N_W(x)$ the set of all neighbors of type W. The {\it dissolution capacity} of the water in the neighborhood of cell $x$ is 
\begin{equation*}
S = 1 - \frac{1}{n_W(x)}\sum\limits_{y \in N_W(x)}  c(y,t).
\end{equation*}
Note that $S \in[0,1]$. We interpret $S$ as  the propensity of the particle at position $x$ to dissolve, and we implement this by allowing the particle at $x$ to begin to  dissolve with probability $S$.  In the CA algorithm a  random number $r$ is drawn uniformly from $[0,1]$, and if $r<S$, the cell's status is set to W and the concentration of the substance that just dissolved is set to 1, while the other concentration remains at 0.
\item {\it Diffusion.}  Cells of type W  have two concentrations associated with them, the concentration of dissolved drug, and the concentration of dissolved excipient.  These concentrations are updated according to the following rule, which is a discretized version of the continuous diffusion equation
\begin{equation*} 
c(x,t_{m+1}) = c(x,t_m) +  \hat{D} \left({ \sum_{y \in N_W(x)} c(y,t_m)  -  n_W(x) c(x,t_m)}  \right) 
\end{equation*}
where $c(x,t_m)$ denotes the concentration in cell $x$ in the updated tablet and $\hat{D}$ is the diffusion rate of the dissolved substance in cell $x$ relative to the expected time that it takes the fastest dissolved substance to diffuse across one grid element.   More precisely,
\begin{equation*} 
\hat{D} = \displaystyle{ \frac{D_i}{ 6 \max \{ D_{drug}, D_{excipient} \} } },
\end{equation*}
and $i$ stands for ``drug" or ``excipient".  This choice ensures that the new value satisfies $0\le\tilde{c}(x)\le1$,  see Appendix \ref{app:diffusion} for more details.
\item {\it Transport.}   Once dissolved drug or excipient reaches the boundary of the tablet, it must diffuse into the surrounding medium.  The concentrations in the grid cells surrounding the tablet are initially set to zero. As molecules diffuse out from the tablet,  the concentrations of drug and 
excipient are updated  in the surrounding layer of water cells.  Since the tablet is assumed to sit in a   large amount of fluid, we can assume that concentrations far from the tablet are relatively constant.  Thus we only model the concentrations in water cells adjacent to the tablet, we call the ``boundary layer".   In order to simulate different stirring mechanisms, the concentrations in cells in the boundary layer are multiplied by a factor 
\mbox{${\tt f}\in[0,1]$} which is provided by the user. 
\end{enumerate}

Since particles dissolve and diffuse at different rates, the real time interval corresponding to one iteration step, $\Delta t$,  is calculated prior to execution of the main loop.  Based on rates provided by the user, $\Delta t$ is set to be the smallest of the four quantities.
\begin{itemize}
\item The average time it takes for a drug respectively excipient particle of mass corresponding to one subcell to dissolve.
\item The average time it takes for a dissolved drug respectively excipient particle to diffuse across one subcell.
\end{itemize}
All other rates are then computed in units of $\Delta t$.  Thus, for example, if the average time it takes for a particle to dissolve is $2$ minutes, and $\Delta t = 0.6\,s$, which is the time it takes for a dissolved drug molecule to diffuse across one grid element, then the dissolution step takes, on average, $\frac{120}{0.6} = 200$ diffusion steps.  We incorporate this variable dissolution rate by decreasing the probability of dissolution by a factor of 200, so that, on average, it will take  $200$ iterations of the algorithm for the particle to start to dissolve.  See Table \ref{parameters} for the default values of {\tt o} and {\tt l}, the average increase in time to dissolution of  drug and excipient particles, respectively. 

\subsection{Release profiles and CA output}
After all  cell states have been updated,  the simulation clock is advanced  it by $\Delta t$, and 
 we calculate the percent drug remaining in the tablet.  We assume that all of the drug in a D cell can be dissolved in a water cell. We will return to this assumption in Section \ref{discussion}.

Let $n_D(t)$ be the number of grid cells in state D at time $t$, so that $n_D(0)$ is the number of 
grid cells initially in state D.  
 Similarly, let ${\bf W}(t)$ denote the set of all wet cells in the tablet at time $t$.
fraction of the initial drug load in the tablet at time $t$ is
\begin{equation*}
M(t)=  \frac{ n_D(t) + \sum_{x \in {\bf W}(t)} c(x,t) }{n_D(0)}.
\end{equation*}
The fraction released at time $t$  is the complement of the fraction of drug in the tablet at that time: 
\begin{equation*}
R(t)=1 - M(t) = 1 -  \frac{ n_D(t) + \sum_{x \in {\bf W}(t)} c(x,t)}{n_D(0)} .
\end{equation*}
The simulation is stopped once a user-specified time has been reached, or when drug is no longer released, i.e.~when $R'(t)$ is (approximately) zero.  The algorithm detects that drug release has stopped when the fraction released  has not changed by more than $10^{-5}$ in 100 consecutive steps. 

\section{Results}\label{results}
The algorithm above has been implemented using \texttt{C++}  in the package \texttt{celldiff} and is available at \cite{codelocation}.  The simulation time and the fraction of drug released up to that time are  exported in ascii format and can easily be read using open source languages such as  \textsc{gnuplot} or  \textsc{scilab}. In addition, the program gives the user the option to export the status of all cells at selected time points as an ascii file, allowing for better insight into the wetting and dissolution processes, see Figure \ref{Fig:Series}.  The state of the simulated tablet suggests explanations for further experimental observations.  For example, Lemaire {\it et al.}~\cite{Lemaire} asked what is the cause of the differences between the initial release of drug and the later release. The CA model describes three phases, namely the uptake of water into the tablet, the dissolution of solid drug and excipient, which causes the formation of pores and finally the diffusion of dissolved drug out of the tablet.  Figure \ref{Fig:Series} shows an interior slice of a tablet $8 \, mm$ in diameter during the initial 8 minutes of  simulation.  The water enters the tablet rapidly, dissolving the soluble particles of drug and excipient.  The interface between solid and dissolved drug is approximately circular in cross-section, and with our choice of parameters,  the radius of the dry region is decreasing at approximately $160\,\mu m\,min^{-1}$. Such  predictions can be tested by immersing dry tablets in a dyed solution for different periods of time. 

In Figure \ref{Fig:baseline} we show the simulated release profile of a drug from a matrix tablet of ${\tt n}=0.8\,cm$ diameter with parameters as in Table \ref{params}. There is strong qualitative agreement with experimental release profiles published in the literature, see e.g.~\cite[Figure 7]{DCDSB}, \cite[Figures 1-4]{Azarmi}, \cite[Figure 2]{Chatterjee2} and \cite[Figures 1-3]{Kuksal}. The agreement here is understood as the presence of the following features.
\begin{itemize}
\item Higher polymer fractions in the powder mixture result in partial release, with the released fraction decreasing as the polymer fraction increases.
\item The early phase of the release is convex, while the later phase is concave.
\item The release occurs over approximately $8\,h$. 
\end{itemize}

\begin{figure}
\hskip -.5in
\begin{minipage}[b]{.3\linewidth}
\includegraphics[scale = .38]{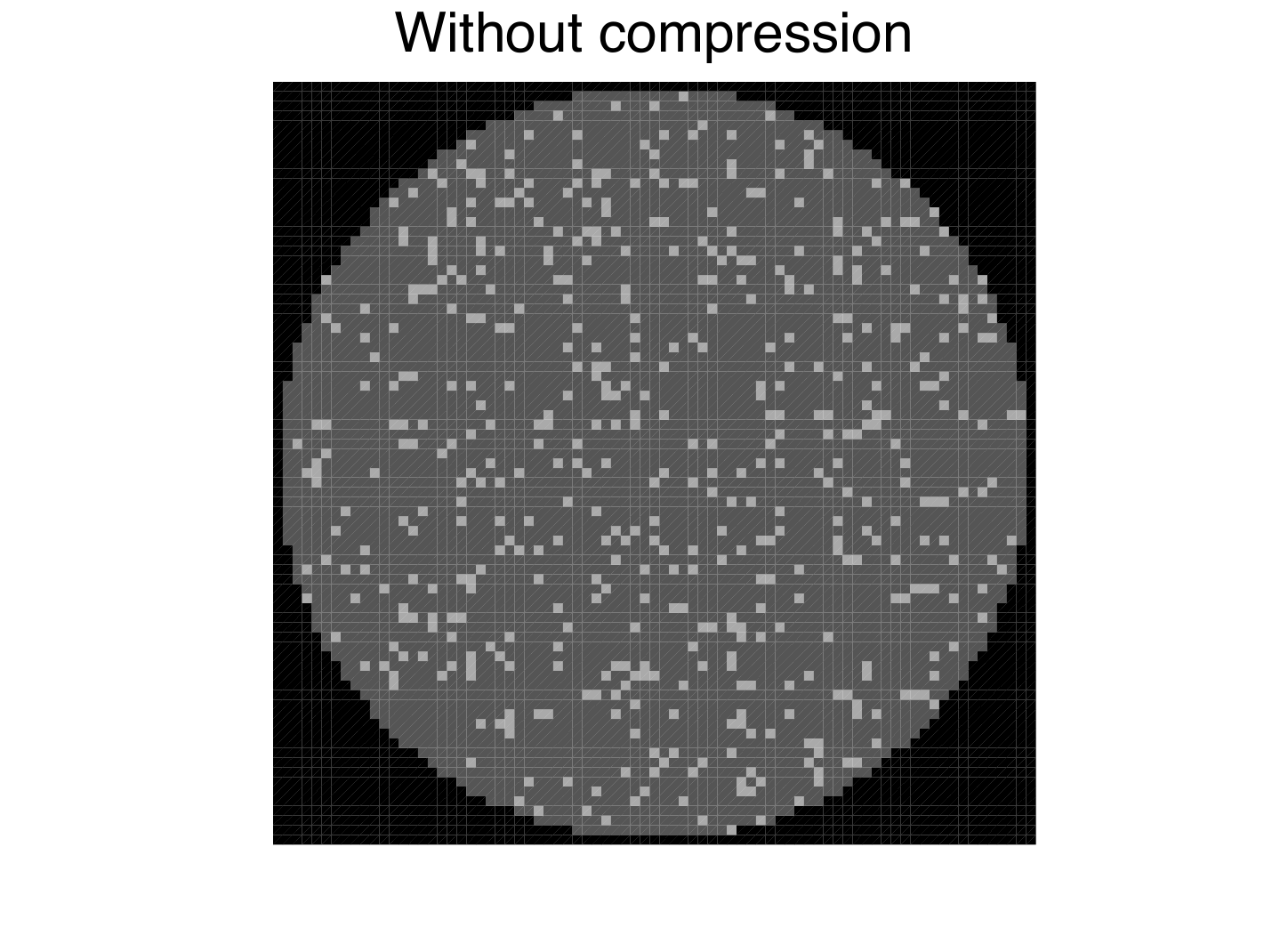} 
\end{minipage}
\begin{minipage}[b]{.3\linewidth}
 \includegraphics[scale = .38]{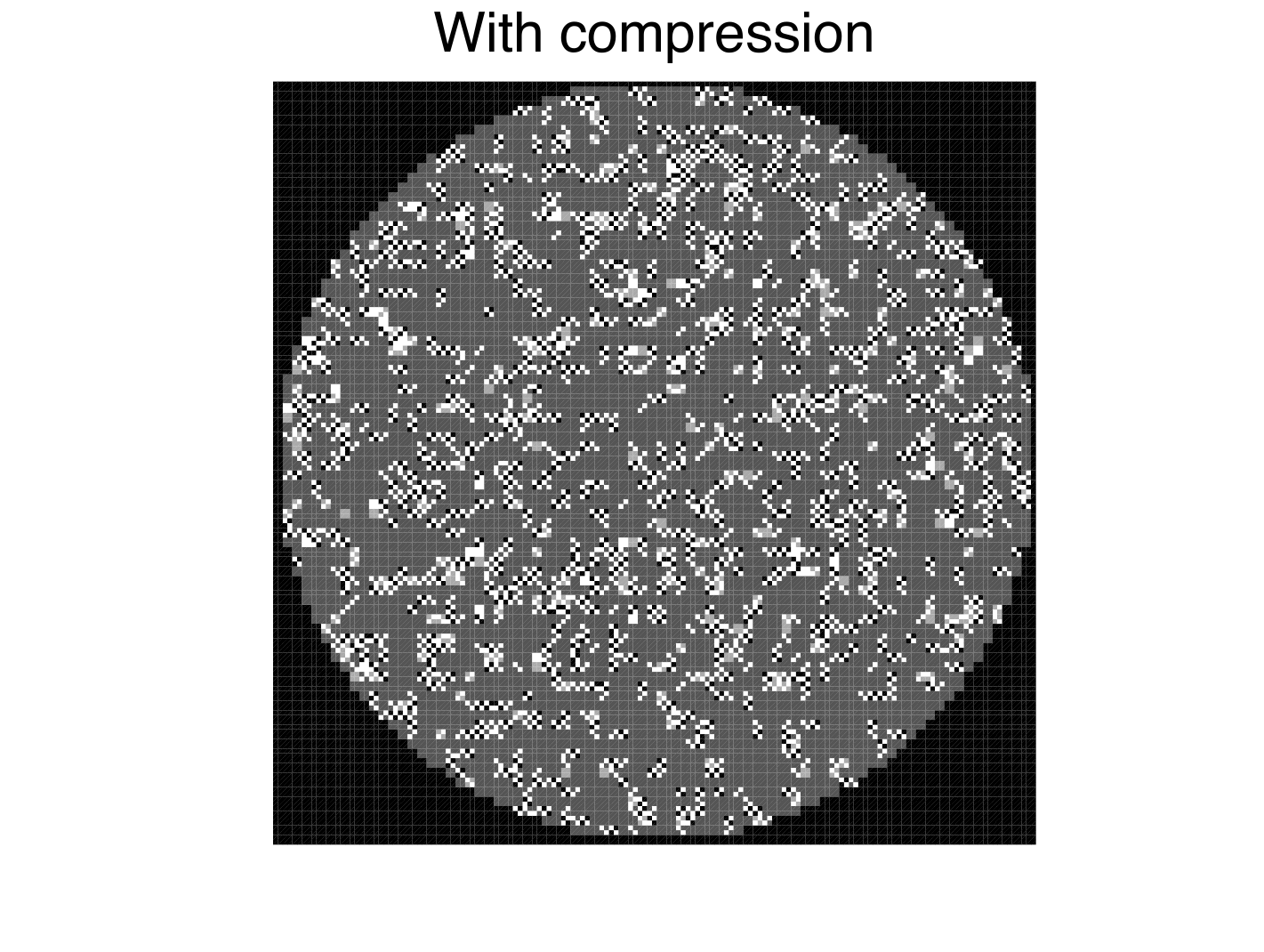} 
 \end{minipage}
 \begin{minipage}[b]{.3\linewidth}
  \includegraphics[scale = .28]{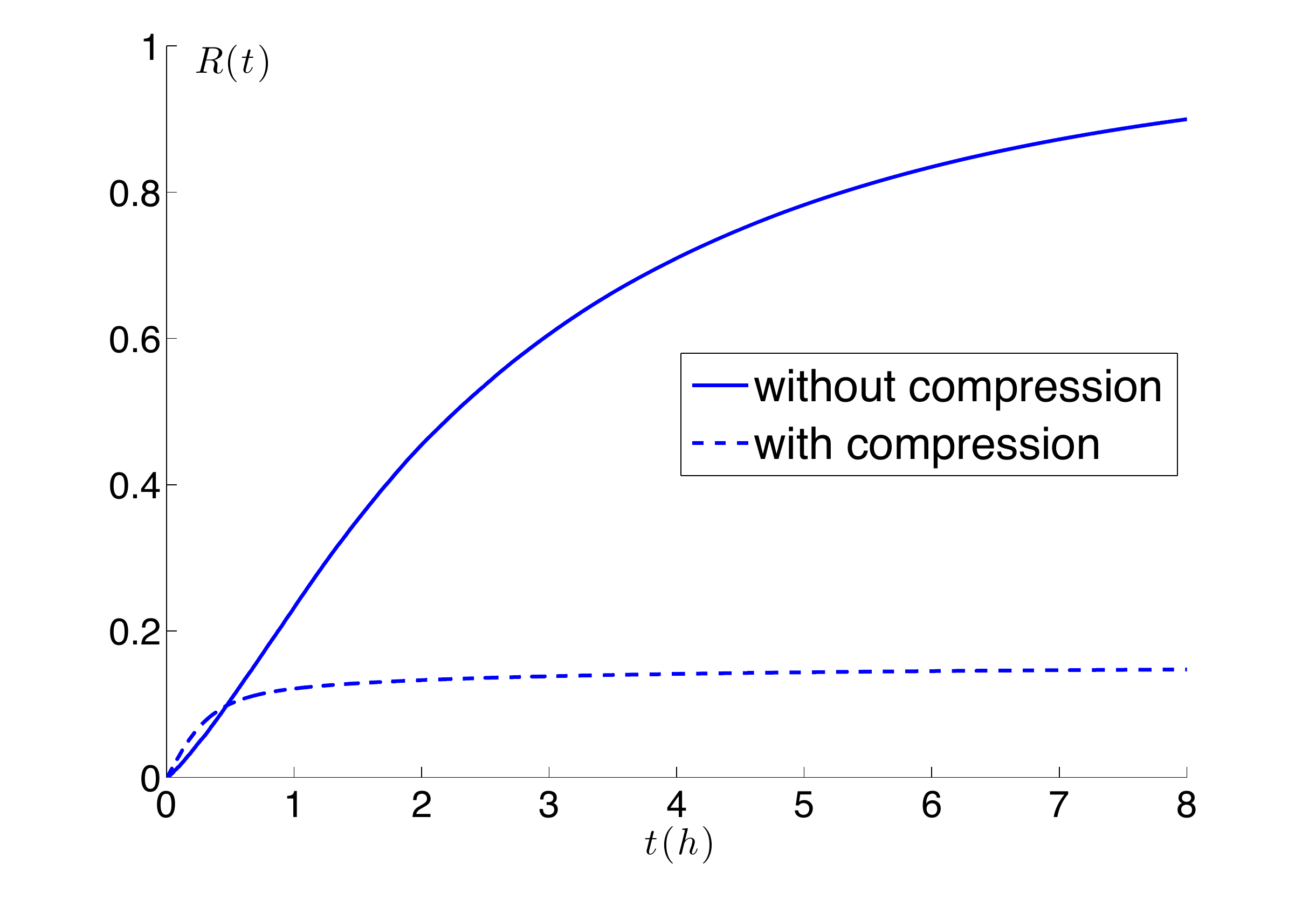}
  \end{minipage}
\caption{Simulated tablets before and after compression.  Left and middle: a slice of the tablet before and after compression.   Polymer cells are lighter, drug and excipient are gray, and boundary, wet or void cells are black.  The lightest cells (middle panel) are cells in which polymer and drug are mixed.   Right: Release curves for a tablet with compression and without compression.   Each tablet contains 50\% polymer.   Other parameters are as given in Table \ref{params}. } \label{Fig:comparecompress}
\end{figure}

\begin{sidewaysfigure}
\includegraphics[scale = .7]{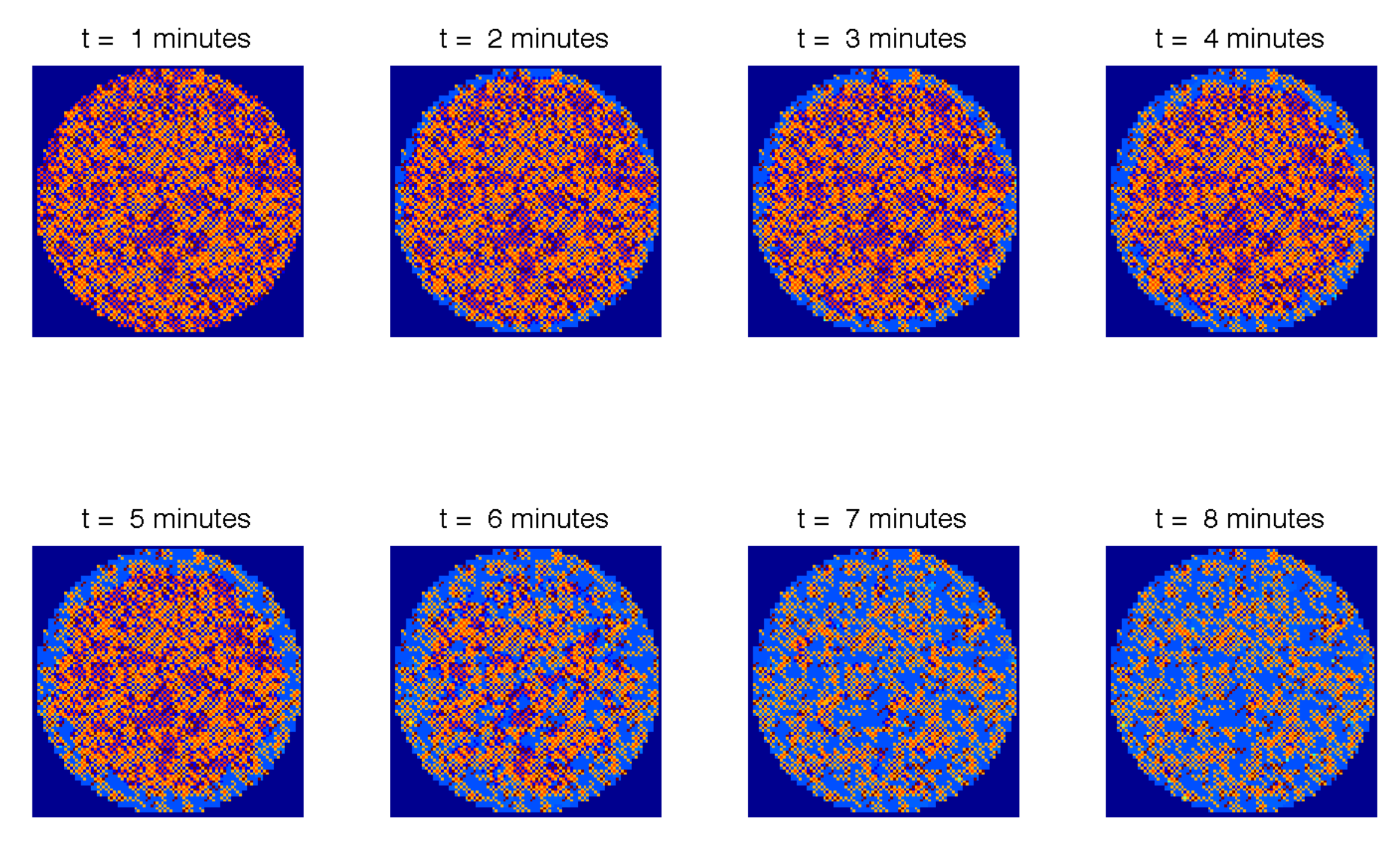}
\caption{A sequence of ``snapshots" of the middle slice of the tablet showing the initial uptake of water and the dissolution of drug and excipient over the first 8 minutes of simulation.  
Light blue colors show the concentration of dissolved drug and excipient.  Red cells represent undissolved drug or excipient particles, yellow cells represent polymer, and dark blue cells represent void or water cells.
The polymer fraction is 30\%;   all  other parameters are as in Table \ref{params}.}
\label{Fig:Series}
\end{sidewaysfigure}

\begin{figure}
\includegraphics[width = 60mm]{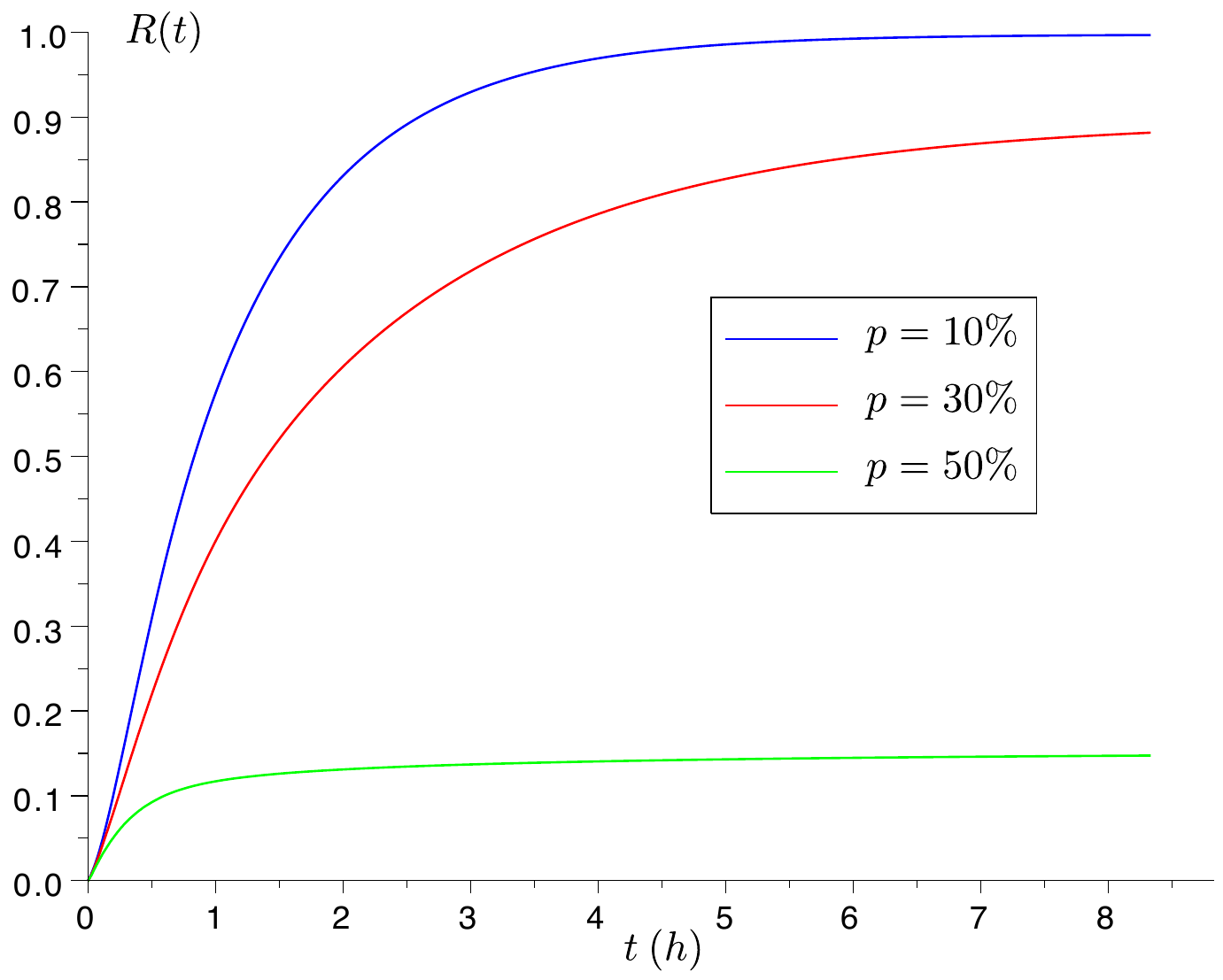}
\caption{Simulation of drug release from a tablet of ${\tt n}=0.8\,cm$ diameter using the parameters in Table \ref{params}.} \label{Fig:baseline}
\end{figure}

\begin{figure}
\includegraphics[width = 60mm]{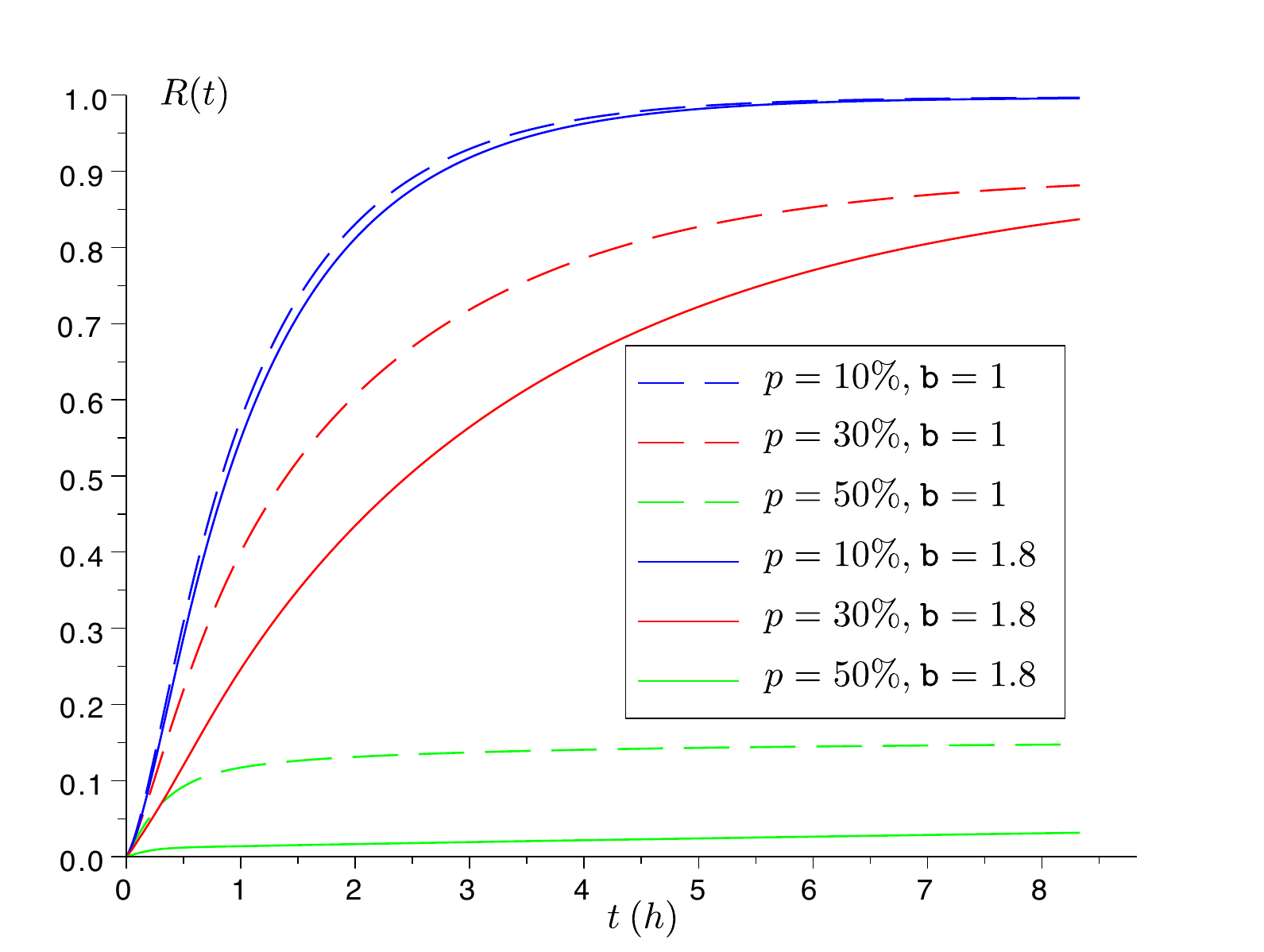}
\includegraphics[width = 60mm]{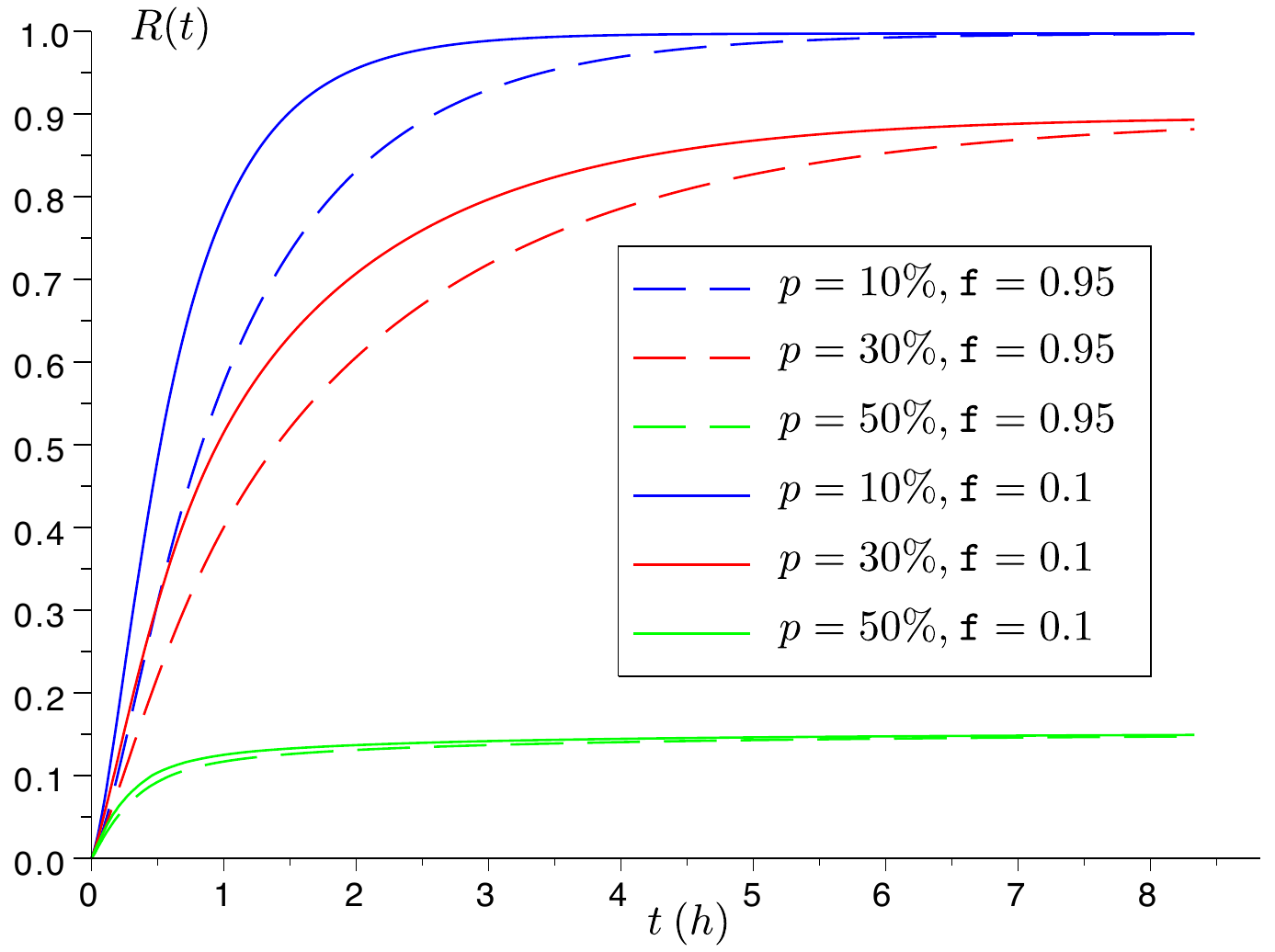}
\includegraphics[width = 60mm]{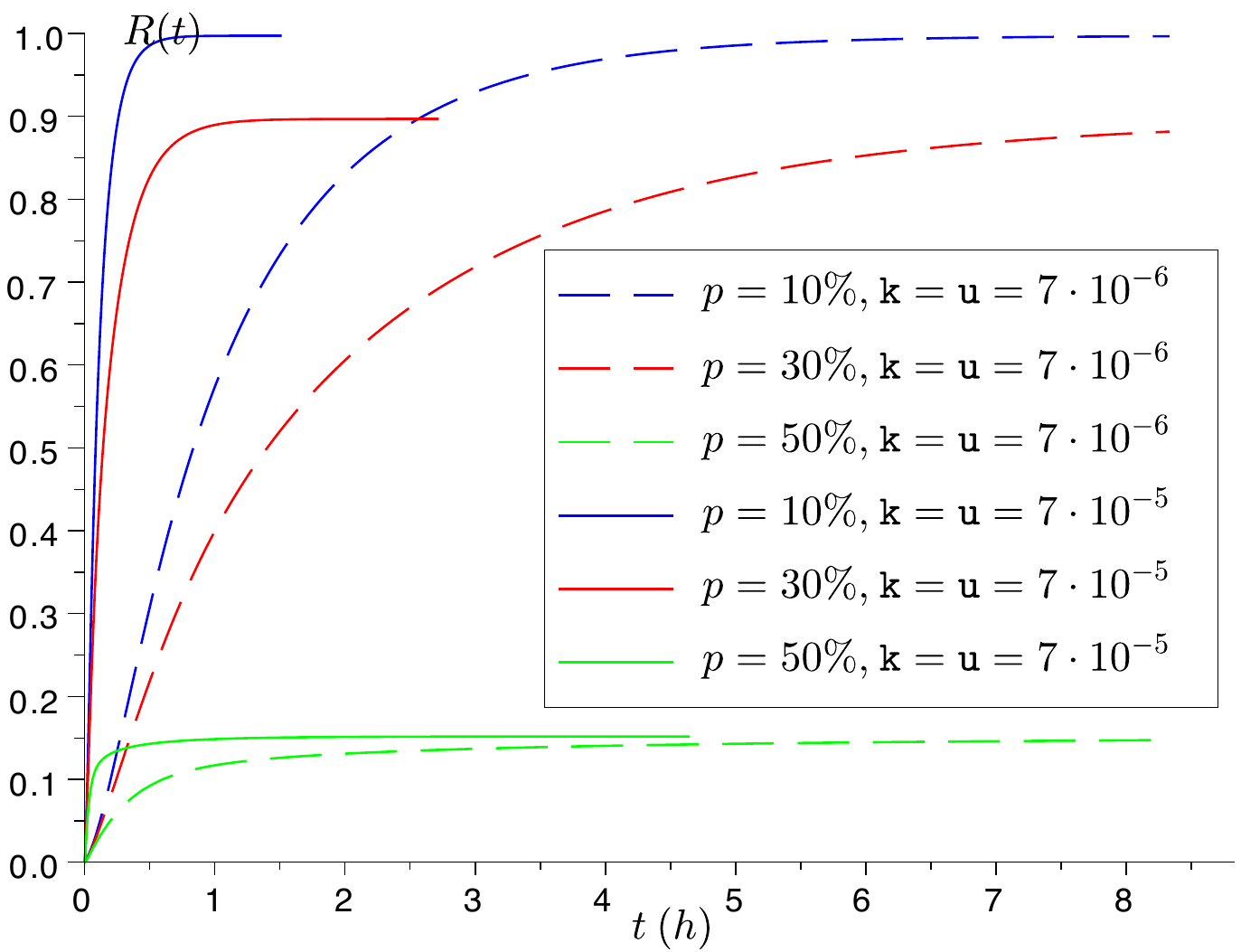}
\includegraphics[width = 60mm]{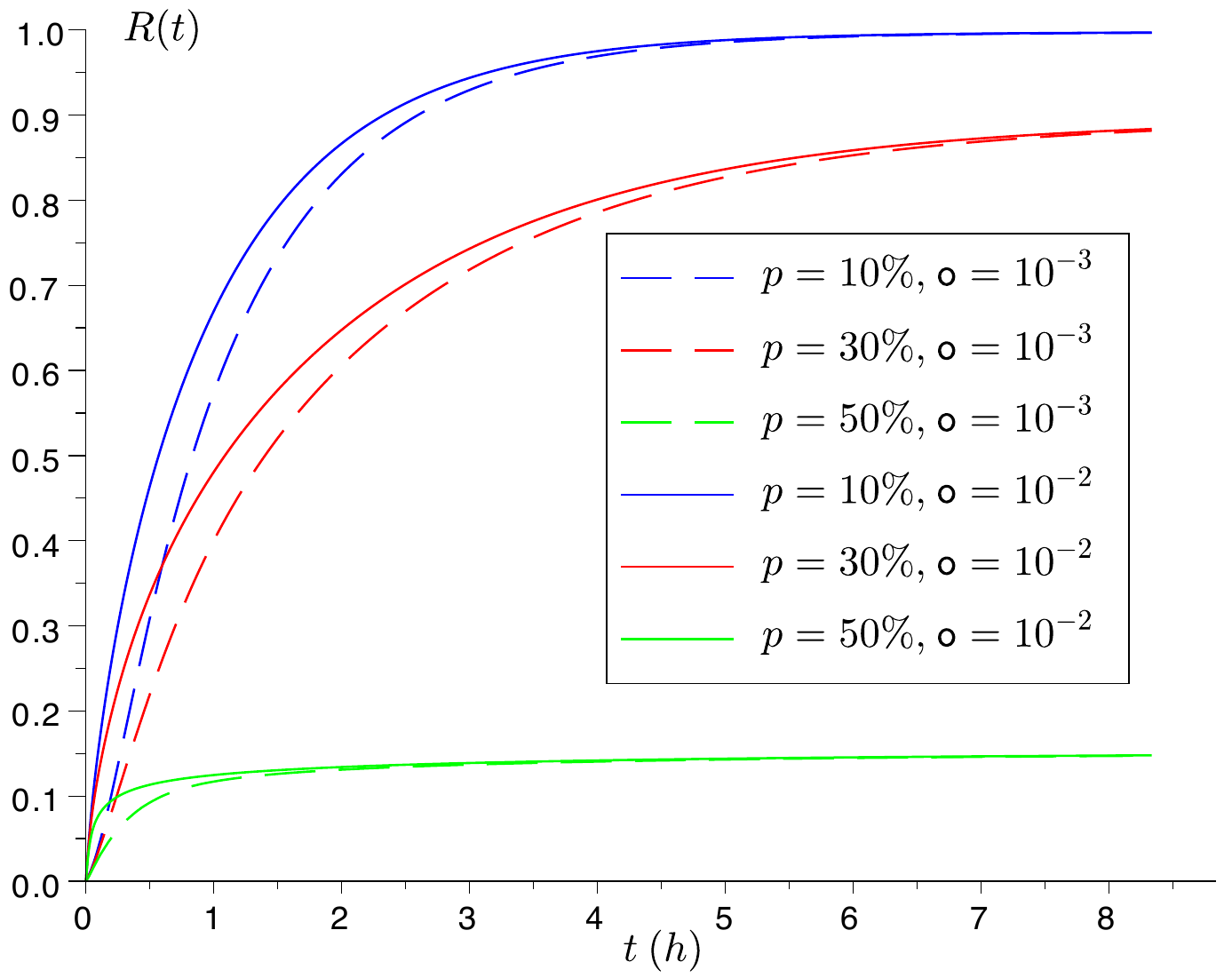}
\caption{Simulation of drug release from a tablet of ${\tt n}=0.8\,cm$ diameter while changing parameters. The dashed lines correspond to the default  parameters in Table \ref{params}.} \label{Fig:change}
\end{figure}

We can study the influence of the individual parameters on the shape of the release profiles by varying them one at a time, see Figure \ref{Fig:change}.  An increase of the 
parameter {\tt b} (Figure \ref{Fig:change}, top left panel), the increased clustering of polymer particles in the outer shell, results in a delayed drug release. In the case of higher polymer fractions, this 
can amount to an almost complete trapping of the drug. The stirring of the surrounding fluid \cite{Viegas} decreases the concentration of dissolved drug outside the tablet and thereby facilitates the diffusion, although this
effect is of somewhat limited size. Clearly, this is of importance mostly for experiments in rotating disk apparatuses and similar devices. An increase of either the diffusion rate {\tt k} or the  
dissolution rate, {\tt o} results in a faster release of the drug. We remark here that the dissolution parameters {\tt o} and {\tt l} affect the rate of  erosion of the tablet, and hence their effect is most noticeable at the initiation of release.  The parameter {\tt k} affects the rate at which drug diffuses from the tablet, and so has a more pronounced effect over the entire release curve.

\section{Discussion}\label{discussion}

From our simulations we gain insight on the role that the individual parameters play and how they influence the drug release profiles. Of course, it has to be borne in mind that some parameters 
 cannot be manipulated in the process of tablet formulation and fabrication, such as dissolution and diffusion rates. Roughly speaking, we can group the parameters into two categories, namely those
that determine the {\it size} of the tablet and those that determine its {\it topology}. In the former category are the diffusion rate {\tt u}, the dissolution rate {\tt o} and, naturally, the tablet diameter {\tt n}.
In the latter category are the width of the melted polymer shell {\tt w} and the imbalance factor {\tt b} that models how much more the polymer particles are clustered in the melting zone at the edge of the tablet. 

\begin{figure}
\includegraphics[width = 60mm]{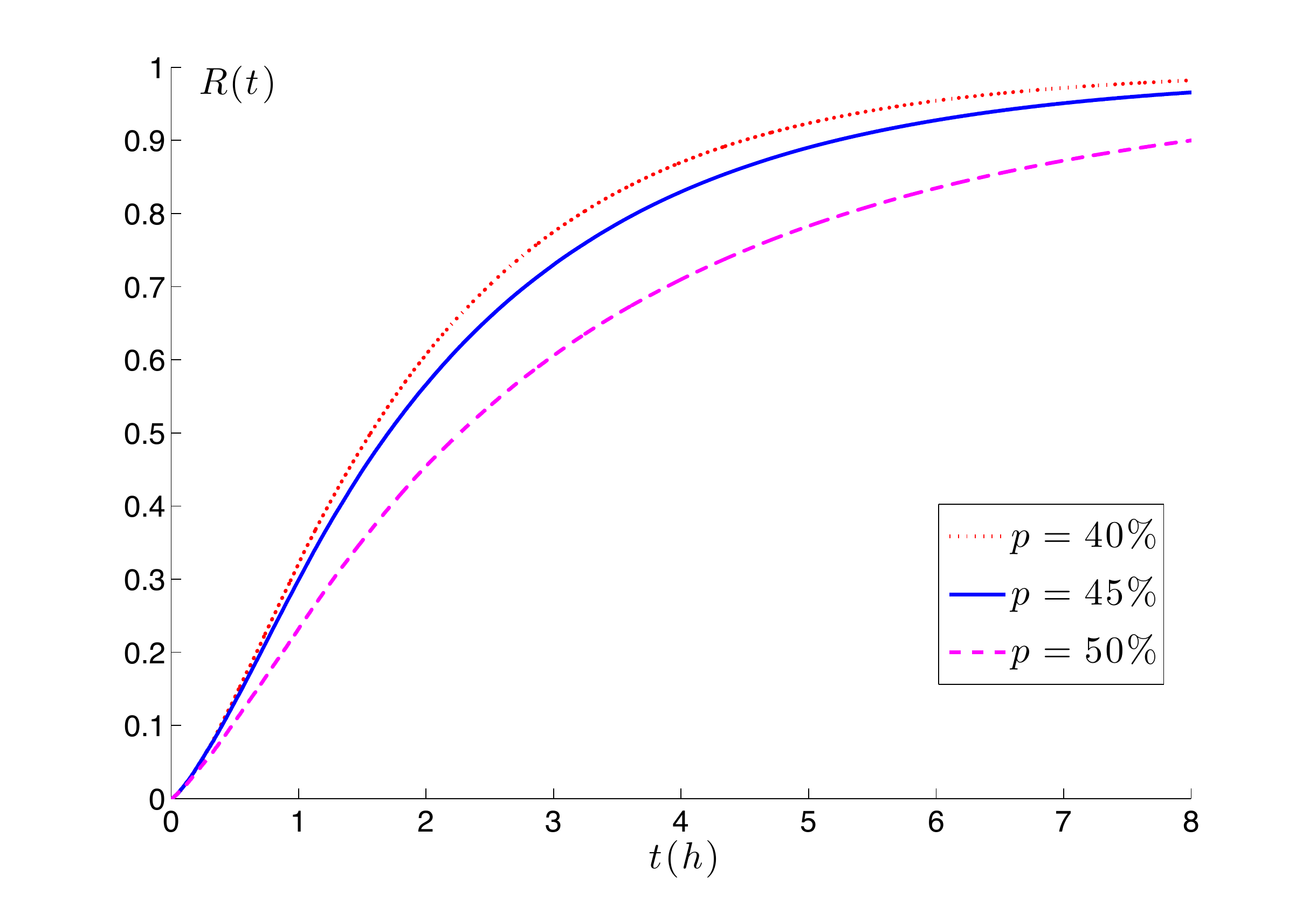}
\includegraphics[width=60mm]{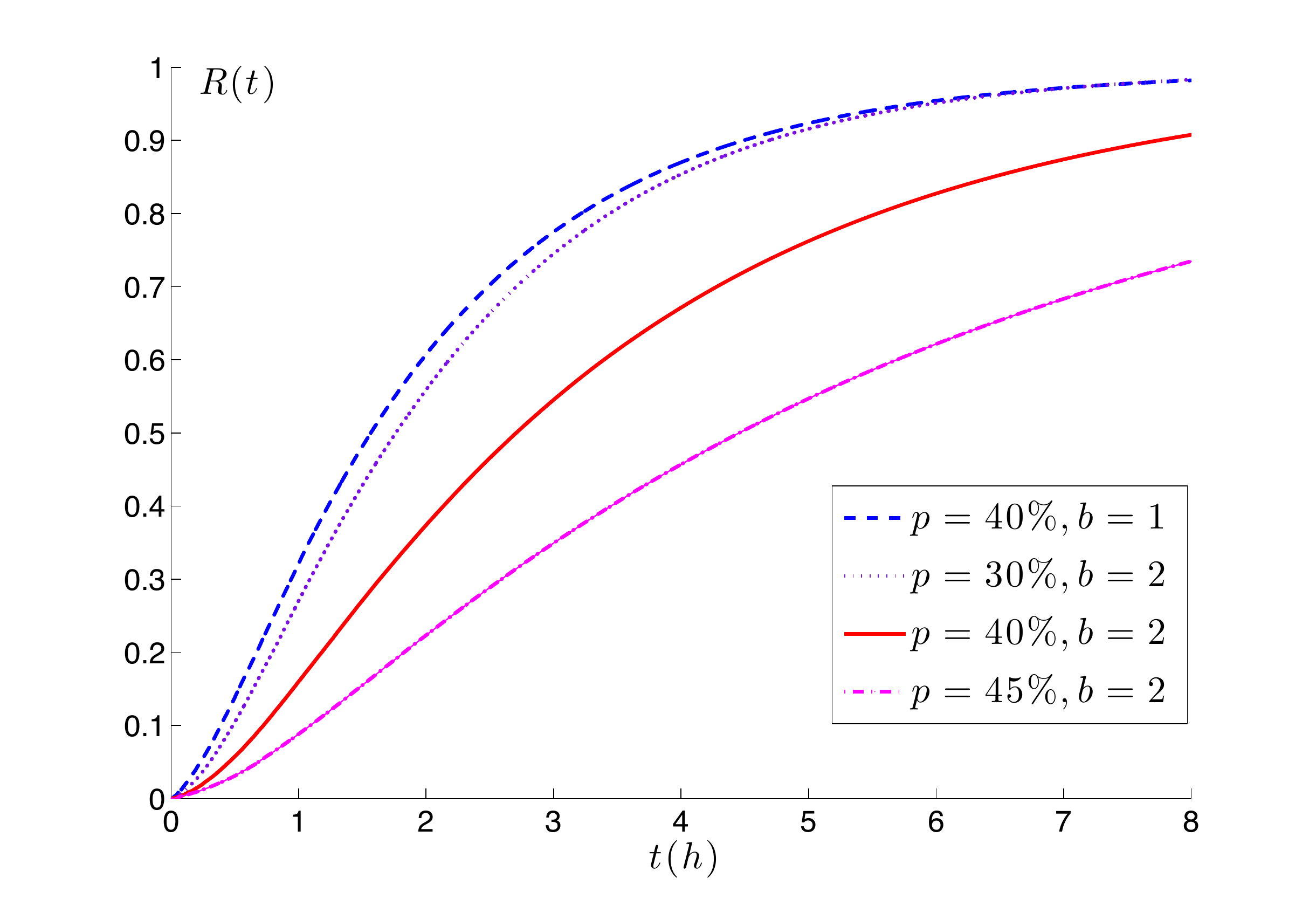}
\caption{(Left) Simulated release curves for tablets containing 40, 45 and 50 \% polymer, modeling no compression or thermal treatment.  The compression flag, {\tt d}, is set to 0.  All other parameters are in Table \ref{params}. (Right) Simulated release curves for tablets containing 30, 40 and 45 \% polymer, with no compression ({\tt d} = 0), with an imposed shell of polymer, {\tt b} = 2.  Comparison is made to a tablet with 40\% polymer and no shell. The release curves are close to the ``ideal", depending on the desired release interval.}\label{Fig:IdealPill}
\end{figure}

{A natural question is the powder composition necessary to produce an ``ideal'' tablet, with a largely linear release profile and a almost complete release of the (expensive) drug.  According to our simulation results and consistent with experimental evidence \cite{Azarmi}, thermal treatment after compression causes drug to be trapped, except at very low polymer fractions.  At these low polymer fractions, however, drug is released very quickly since most of the tablet erodes.  Therefore, in order to get complete release of the drug over 8 hours, with a nearly linear release profile, the model suggests using a polymer fraction between 40 and 45\%, with as little deformation and fusing of the polymer particles as possible.  We illustrate this in Figure \ref{Fig:IdealPill}, which shows simulations in which the compression sub-routine is disabled (flag {\tt d} is set to 0).   Note that in this model we assume brittle drug and excipient, such as indomethacin and lactose, and a deformable, non-soluble, non-permeable polymer.  If the properties of the particles are changed, then the model must be modified accordingly.}   {For a release profile close to the ideal it is desirable to put a shell of polymer on the tablet without completely enclosing the interior.  In our model, we set {\tt b} = 2, but we do not apply compression or heating.  With $30 \%$ polymer, we get a close to ideal release over 8 hours, and higher polymer concentrations extend the release profile over longer time intervals.  See Figure \ref{Fig:IdealPill} (right panel).}

Of the possible  drug release mechanisms, our model mimics the transport of drug molecules through water filled pores. As time passes, degradation of the excipient and drug particles occurs and many crevices and channels are formed. The diffusion through these pores is highly dependent on the polymer structure and consequently also dependent on the processes that alter pore formation and closure \cite{Fredenberg}. As it can be seen in Figure \ref{Fig:comparecompress}, the spreading and fusing of polymer particles after compression and thermal treatment changes the tablet structure and modifies the tortuosity, pore formation and the resulting release curves. This structural change is crucial to reproduce the trapping of drug at high polymer concentrations observed in experimental data. Without it, the release obtained at high polymer concentrations did not match that of the experimental data. 
The initial burst release observed in Figure \ref{Fig:baseline} for all polymer densities, can be attributed to drug particles easily accessible by hydration on the surface ~\cite{Fredenberg,Wang}. Thus, to match the very low rate of drug release present at high curing temperatures and high polymer content, we implement the formation of a polymer shell on the surface of the tablet. As described in \cite{Azarmi}, the effect of thermal treating causes a better coalescence of polymer particles with a decreased porosity and a smoother surface.
By altering the morphology of the outer layer, we successfully control the permeability and eliminate the initial burst, providing a more faithful picture of the experimental results.

To include other release mechanisms such as diffusion through the polymer or degradation of the polymer in our mathematical model, it will be necessary to describe the diffusion coefficient as a function of other parameters. The diffusion rate through a polymer is highly dependent on its physical state.  For some polymers, there is a  glass temperature transition in which the polymer changes from brittle to rubber-like. In this case,  the diffusion rate can change by several orders of magnitude \cite{Fredenberg}.  For biodegradable polymers, modeling the dissolution of the polymer and the corresponding matrix degradation might require a non-constant diffusion coefficient that varies in proportion to the local fraction of polymer in the tablet \cite{Wada}. 

The role of the excipient in our model is merely that of a space filler, which is not to say that it is not important. Dissolving excipient opens new channels through which  water can  reach the drug, and through which the drug can diffuse out of the tablet.   It should be stressed that our assumption that drug and excipient dissolve independently of each other in water is a simplification. A water cell  whose water is saturated with respect to excipient should have
a very small capacity to dissolve drug and vice versa. Furthermore, the current assumption that the entire mass of drug in one grid cell can  be dissolved in the amount of water in a grid cell of the same volume is another simplification. A more realistic implementation of the model would incorporate a gradual dissolution process, so that only a fraction of the drug in a D cell is dissolved in one simulation step.

One simulation representing 8 hours with the baseline parameters given in Table \ref{params} typically takes one hour on a MacPro workstation (6-Core Intel Xeon
  at	$2.93\, GHz$).  The algorithm is easily parallelizable, which is one of the positive features of a CA model.    This is work in progress.

\begin{appendix}
\section{Sample call and parameter values}\label{parameters}

A sample call of the code is as follows: user-supplied arguments are preceded by a hyphen, and successive arguments are separated by spaces.
For example, the call
\begin{verbatim}
./celldiff -n 0.8  -p 0.3 -c30000 -t100 -sTabletData.txt
\end{verbatim}
calls the program simulating a tablet with diameter ${\tt n}= 0.8\,cm$, made up of 30\% polymer and 10\% drug (the default).  It simulates the evolution of the tablet over $3\cdot10^4\, s$  and exports the full state data every 100 time steps, storing it in the text file {\tt TabletData.txt}.  
Table \ref{params}  lists  the model parameters and their default values in the program, {\tt celldiff}. 
\begin{table}
\begin{tabular}{ |c|c|p{1.5in}| c| c|} \hline
Name & User Input & Meaning & Default Value & Units/Range \\ \hline 
diameter & {\tt n} & diameter of the computational domain &  $0.8$ & $cm$ \\ \hline
maxtime & {\tt c} & maximum time to simulate before halting & {$30000$}  &	$s$	\\ \hline
polymer & {\tt p} & polymer concentration &0.3 & $[0,1]$	\\ \hline
drug	& {\tt g} & drug concentration &0.1 & $[0,1]$	\\ \hline	
tablet height & {\tt h} & height of tablet as a ratio of diameter &0.23& $[0,1]$ 	\\ \hline
release file & {\tt r} & filename for exporting release curve & time stamped name & \\ \hline	
state file & {\tt s} & filename for exporting full state data & time stamped name &\\ \hline	
stateperiod & {\tt t} & how often to export full state data &0  (never)& \\ \hline	
compress & {\tt d}& flag to indicate whether or not compression and thermal treatment is implemented& 1 & $1$ or $0$ \\ \hline	
seed	& {\tt e}& random number seed & 47 & $\mathbb{N}$ \\ \hline
asciiperiod	& {\tt a} &	frame period in iterations for drawing ascii animation&1&  $\mathbb{N}$ \\ \hline
drugdissprob & {\tt o} & dissolution probability scale for drug &{$10^{-3}$}& \\ \hline
exdissprob	& {\tt l} & dissolution probability scale for excipient &{$10^{-3}$}&	\\ \hline
polyshellbalance & {\tt b}& distribution of polymer in the shell &1& $\ge 1$ \\ \hline
{removal rate} &{\tt f}& removal factor for concentrations in boundary cells & {$0.95$}& $[0,1]$ \\ \hline
nographics	& {\tt x} & set  $>0$ to run in non-interactive, text-only mode& 1&	\\ \hline
drugdiffrate & {\tt u} &	physical rate of drug diffusion & $7\cdot10^{-6}$& $\displaystyle{\frac{cm}{s^2}}$ \\ \hline
exdiffrate & {\tt k} & physical rate of excipient diffusion&	$7\cdot10^{-6}$ & $\displaystyle{\frac{cm}{s^2}}$ \\ \hline
cellsize & {\tt y}& physical size of a cell & $0.01$& $cm$\\ \hline
polyshellwidth & {\tt w} & width of the polymer shell (has no effect when $b = 1$, the default)  &1& cell size  \\ \hline													
\end{tabular}
\caption{Default values of the parameters in {\tt celldiff}. }\label{params}
\end{table}

\section{Derivation of the diffusion update rule}\label{app:diffusion}

Let $c(x,t)$ be the concentration of a substance at a point $x \in {\mathbb R}^3$ at time $t \in {\mathbb R}$.  The diffusion equation in three dimensions is  
\begin{equation*}
\frac{\partial c}{\partial t}  = D \Delta c
\end{equation*}
where $\Delta$ denotes the usual Laplace operator in three dimensions and $D$ is the diffusion constant (with units Length$^2$/Time).  After discretization of space on a cubic lattice with lattice constant $L$ and of time, so that  $x = (i,j,k)$ and  $t_m = m \Delta t$,
the discretized version of the diffusion equation becomes
\begin{equation} \label{Eq:discretediff}
 c(x,t_{m+1}) - c(x,t_m) = \frac{ D \Delta t }{L^2} \left({ \sum_{y  \in N(x)} c(y,t_m)  - 6 c({x,t_m}) }\right),
\end{equation}
where $N({x})$ is  the von Neumann neighborhood of ${x}$.
In the CA model, we choose $\Delta t$ to be the expected time it would take for a dissolved particle to diffuse across one grid element, of side length $L$.  Since the variance of the diffusion process, $X(t)$,  is given by 
$E[X^2] =  6 D t $,
we choose the time step, $\Delta t$ so that $L^2 = 6 D \Delta t$ and hence $\Delta t = \frac{L^2}{6 D}$, 
where $D = \max \{ D_{drug}, D_{excipient} \}$ is  the maximum of the diffusion rates of the drug and excipient.  Substituting this value of $\Delta t$ into the discretized diffusion equation, equation \eqref{Eq:discretediff}, and rearranging gives 
\begin{equation}\label{Eq:diffusionupdate}
c(x,t_{m+1}) = c(x,t_m) + \hat{D}  \left({ \sum_{{y}  \in N({x})} c(y,t_m)  - 6 c(x,t_m) }\right) 
\end{equation}
 where $\hat{D} = \displaystyle{ \frac{D_i}{ 6 \max \{ D_{drug}, D_{excipient} \} } }$, where $i$ is ``drug" or ``excipient", depending on the concentration that is updated.
 In this model, certain neighbors might contain solid particles of drug or excipient, or polymer.  We impose zero flux conditions on the boundaries to these solid neighbors, so that only terms involving neighboring cells in the ``wet" (W) state are counted in the expression in equation \eqref{Eq:diffusionupdate}.  This gives the diffusion update rule described above.
\end{appendix}

\section*{Acknowledgments}
This work has been supported by the grant ``Collaborative Research: Predicting the Release Kinetics of Matrix Tablets'' (DMS 1016214 to Peter Hinow and DMS
1016136 to Ami Radunskaya) of the National Science Foundation of the United States of America. We thank Drs.~Boris B\"aumer (Department of Mathematics and Statistics), Lipika Chatterjee and Ian Tucker (School of Pharmacy) at the University of Otago in Dunedin, New Zealand, for inspiring discussions. 
\bibliographystyle{apsrev4-1}
\bibliography{MatrixTablets}

\begin{thebibliography}{18}%
\makeatletter
\providecommand \@ifxundefined [1]{%
 \@ifx{#1\undefined}
}%
\providecommand \@ifnum [1]{%
 \ifnum #1\expandafter \@firstoftwo
 \else \expandafter \@secondoftwo
 \fi
}%
\providecommand \@ifx [1]{%
 \ifx #1\expandafter \@firstoftwo
 \else \expandafter \@secondoftwo
 \fi
}%
\providecommand \natexlab [1]{#1}%
\providecommand \enquote  [1]{``#1''}%
\providecommand \bibnamefont  [1]{#1}%
\providecommand \bibfnamefont [1]{#1}%
\providecommand \citenamefont [1]{#1}%
\providecommand \href@noop [0]{\@secondoftwo}%
\providecommand \href [0]{\begingroup \@sanitize@url \@href}%
\providecommand \@href[1]{\@@startlink{#1}\@@href}%
\providecommand \@@href[1]{\endgroup#1\@@endlink}%
\providecommand \@sanitize@url [0]{\catcode `\\12\catcode `\$12\catcode
  `\&12\catcode `\#12\catcode `\^12\catcode `\_12\catcode `\%12\relax}%
\providecommand \@@startlink[1]{}%
\providecommand \@@endlink[0]{}%
\providecommand \url  [0]{\begingroup\@sanitize@url \@url }%
\providecommand \@url [1]{\endgroup\@href {#1}{\urlprefix }}%
\providecommand \urlprefix  [0]{URL }%
\providecommand \Eprint [0]{\href }%
\@ifxundefined \urlstyle {%
  \providecommand \doi  [0]{\begingroup \@sanitize@url \@doi}%
  \providecommand \@doi [1]{\endgroup \@@startlink {\doibase
  #1}doi:\discretionary {}{}{}#1\@@endlink }%
}{%
  \providecommand \doi  [0]{doi:\discretionary{}{}{}\begingroup
  \urlstyle{rm}\Url }%
}%
\providecommand \doibase [0]{http://dx.doi.org/}%
\providecommand \Doi [0]{\begingroup \@sanitize@url \@Doi }%
\providecommand \@Doi  [1]{\endgroup\@@startlink{\doibase#1}\@@Doi}%
\providecommand \@@Doi [1]{#1\@@endlink}%
\providecommand \selectlanguage [0]{\@gobble}%
\providecommand \bibinfo  [0]{\@secondoftwo}%
\providecommand \bibfield  [0]{\@secondoftwo}%
\providecommand \translation [1]{[#1]}%
\providecommand \BibitemOpen [0]{}%
\providecommand \bibitemStop [0]{}%
\providecommand \bibitemNoStop [0]{.\EOS\space}%
\providecommand \EOS [0]{\spacefactor3000\relax}%
\providecommand \BibitemShut  [1]{\csname bibitem#1\endcsname}%
\bibitem [{\citenamefont {Fredenberg}\ \emph {et~al.}(2011)\citenamefont
  {Fredenberg}, \citenamefont {Wahlgren}, \citenamefont {Reslow},\ and\
  \citenamefont {Axelsson}}]{Fredenberg}%
  \BibitemOpen
  \bibfield  {author} {\bibinfo {author} {\bibfnamefont {S.}~\bibnamefont
  {Fredenberg}}, \bibinfo {author} {\bibfnamefont {M.}~\bibnamefont
  {Wahlgren}}, \bibinfo {author} {\bibfnamefont {M.}~\bibnamefont {Reslow}}, \
  and\ \bibinfo {author} {\bibfnamefont {A.}~\bibnamefont {Axelsson}},\
  }\href@noop {} {\bibfield  {journal} {\bibinfo  {journal} {Int. J. Pharm.},\
  }\textbf {\bibinfo {volume} {415}},\ \bibinfo {pages} {34} (\bibinfo {year}
  {2011})}\BibitemShut {NoStop}%
\bibitem [{\citenamefont {Zuleger}\ and\ \citenamefont
  {Lippold}(2001)}]{Zuleger}%
  \BibitemOpen
  \bibfield  {author} {\bibinfo {author} {\bibfnamefont {S.}~\bibnamefont
  {Zuleger}}\ and\ \bibinfo {author} {\bibfnamefont {B.~C.}\ \bibnamefont
  {Lippold}},\ }\href@noop {} {\bibfield  {journal} {\bibinfo  {journal} {Int.
  J. Pharm.},\ }\textbf {\bibinfo {volume} {217}},\ \bibinfo {pages} {139}
  (\bibinfo {year} {2001})}\BibitemShut {NoStop}%
\bibitem [{\citenamefont {Casault}\ and\ \citenamefont
  {Slater}(2008)}]{Casault_PhysA}%
  \BibitemOpen
  \bibfield  {author} {\bibinfo {author} {\bibfnamefont {S.}~\bibnamefont
  {Casault}}\ and\ \bibinfo {author} {\bibfnamefont {G.~W.}\ \bibnamefont
  {Slater}},\ }\href@noop {} {\bibfield  {journal} {\bibinfo  {journal}
  {Physica A},\ }\textbf {\bibinfo {volume} {387}},\ \bibinfo {pages} {5387}
  (\bibinfo {year} {2008})}\BibitemShut {NoStop}%
\bibitem [{\citenamefont {Lemaire}\ \emph {et~al.}(2003)\citenamefont
  {Lemaire}, \citenamefont {B\'elair},\ and\ \citenamefont
  {Hildgen}}]{Lemaire}%
  \BibitemOpen
  \bibfield  {author} {\bibinfo {author} {\bibfnamefont {V.}~\bibnamefont
  {Lemaire}}, \bibinfo {author} {\bibfnamefont {J.}~\bibnamefont {B\'elair}}, \
  and\ \bibinfo {author} {\bibfnamefont {P.}~\bibnamefont {Hildgen}},\
  }\href@noop {} {\bibfield  {journal} {\bibinfo  {journal} {Int. J. Pharm.},\
  }\textbf {\bibinfo {volume} {{\bf 258}}},\ \bibinfo {pages} {95} (\bibinfo
  {year} {2003})}\BibitemShut {NoStop}%
\bibitem [{\citenamefont {Villalobos}\ \emph {et~al.}(2006)\citenamefont
  {Villalobos}, \citenamefont {Cordero}, \citenamefont {Vidales},\ and\
  \citenamefont {Dom\'inguez}}]{Villalobos_PhysA}%
  \BibitemOpen
  \bibfield  {author} {\bibinfo {author} {\bibfnamefont {R.}~\bibnamefont
  {Villalobos}}, \bibinfo {author} {\bibfnamefont {S.}~\bibnamefont {Cordero}},
  \bibinfo {author} {\bibfnamefont {A.~M.}\ \bibnamefont {Vidales}}, \ and\
  \bibinfo {author} {\bibfnamefont {A.}~\bibnamefont {Dom\'inguez}},\
  }\href@noop {} {\bibfield  {journal} {\bibinfo  {journal} {Physica A},\
  }\textbf {\bibinfo {volume} {367}},\ \bibinfo {pages} {305} (\bibinfo {year}
  {2006})}\BibitemShut {NoStop}%
\bibitem [{\citenamefont {Siepmann}\ and\ \citenamefont
  {Peppas}(2000)}]{Siepmann}%
  \BibitemOpen
  \bibfield  {author} {\bibinfo {author} {\bibfnamefont {J.}~\bibnamefont
  {Siepmann}}\ and\ \bibinfo {author} {\bibfnamefont {N.~A.}\ \bibnamefont
  {Peppas}},\ }\href@noop {} {\bibfield  {journal} {\bibinfo  {journal} {Pharm.
  Res.},\ }\textbf {\bibinfo {volume} {{\bf 17}}},\ \bibinfo {pages} {1290}
  (\bibinfo {year} {2000})}\BibitemShut {NoStop}%
\bibitem [{\citenamefont {Baeumer}\ \emph {et~al.}(2009)\citenamefont
  {Baeumer}, \citenamefont {Chatterjee}, \citenamefont {Hinow}, \citenamefont
  {Rades}, \citenamefont {Radunskaya},\ and\ \citenamefont {Tucker}}]{DCDSB}%
  \BibitemOpen
  \bibfield  {author} {\bibinfo {author} {\bibfnamefont {B.}~\bibnamefont
  {Baeumer}}, \bibinfo {author} {\bibfnamefont {L.}~\bibnamefont {Chatterjee}},
  \bibinfo {author} {\bibfnamefont {P.}~\bibnamefont {Hinow}}, \bibinfo
  {author} {\bibfnamefont {T.}~\bibnamefont {Rades}}, \bibinfo {author}
  {\bibfnamefont {A.}~\bibnamefont {Radunskaya}}, \ and\ \bibinfo {author}
  {\bibfnamefont {I.~G.}\ \bibnamefont {Tucker}},\ }\href@noop {} {\bibfield
  {journal} {\bibinfo  {journal} {Discr. Contin. Dyn. Sys. B},\ }\textbf
  {\bibinfo {volume} {12}},\ \bibinfo {pages} {261} (\bibinfo {year}
  {2009})}\BibitemShut {NoStop}%
\bibitem [{\citenamefont {Viegas}\ \emph {et~al.}(2001)\citenamefont {Viegas},
  \citenamefont {Curatella}, \citenamefont {Van~Winkle},\ and\ \citenamefont
  {Brinker}}]{Viegas}%
  \BibitemOpen
  \bibfield  {author} {\bibinfo {author} {\bibfnamefont {T.~X.}\ \bibnamefont
  {Viegas}}, \bibinfo {author} {\bibfnamefont {R.~U.}\ \bibnamefont
  {Curatella}}, \bibinfo {author} {\bibfnamefont {L.~L.}\ \bibnamefont
  {Van~Winkle}}, \ and\ \bibinfo {author} {\bibfnamefont {G.}~\bibnamefont
  {Brinker}},\ }\href@noop {} {\bibfield  {journal} {\bibinfo  {journal}
  {Pharm. Technol.},\ }\textbf {\bibinfo {volume} {25}},\ \bibinfo {pages} {44}
  (\bibinfo {year} {2001})}\BibitemShut {NoStop}%
\bibitem [{\citenamefont {Nowak}\ and\ \citenamefont
  {Schadschneider}(2012)}]{Nowak_PRE85}%
  \BibitemOpen
  \bibfield  {author} {\bibinfo {author} {\bibfnamefont {S.}~\bibnamefont
  {Nowak}}\ and\ \bibinfo {author} {\bibfnamefont {A.}~\bibnamefont
  {Schadschneider}},\ }\href@noop {} {\bibfield  {journal} {\bibinfo  {journal}
  {Phys. Rev. E},\ }\textbf {\bibinfo {volume} {85}},\ \bibinfo {pages}
  {066128} (\bibinfo {year} {2012})}\BibitemShut {NoStop}%
\bibitem [{\citenamefont {Goltsev}\ \emph {et~al.}(2010)\citenamefont
  {Goltsev}, \citenamefont {de~Abreu}, \citenamefont {Dorogovtsev},\ and\
  \citenamefont {Mendes}}]{Goltsev_PRE81}%
  \BibitemOpen
  \bibfield  {author} {\bibinfo {author} {\bibfnamefont {A.~V.}\ \bibnamefont
  {Goltsev}}, \bibinfo {author} {\bibfnamefont {F.~V.}\ \bibnamefont
  {de~Abreu}}, \bibinfo {author} {\bibfnamefont {S.~N.}\ \bibnamefont
  {Dorogovtsev}}, \ and\ \bibinfo {author} {\bibfnamefont {J.~F.~F.}\
  \bibnamefont {Mendes}},\ }\href@noop {} {\bibfield  {journal} {\bibinfo
  {journal} {Phys. Rev. E},\ }\textbf {\bibinfo {volume} {81}},\ \bibinfo
  {pages} {061921} (\bibinfo {year} {2010})}\BibitemShut {NoStop}%
\bibitem [{\citenamefont {Kavousanakis}\ \emph {et~al.}(2012)\citenamefont
  {Kavousanakis}, \citenamefont {Liu}, \citenamefont {Boudouvis}, \citenamefont
  {Lowengrub},\ and\ \citenamefont {Kevrekidis}}]{Kavousanakis_PRE85}%
  \BibitemOpen
  \bibfield  {author} {\bibinfo {author} {\bibfnamefont {M.~E.}\ \bibnamefont
  {Kavousanakis}}, \bibinfo {author} {\bibfnamefont {P.}~\bibnamefont {Liu}},
  \bibinfo {author} {\bibfnamefont {A.~G.}\ \bibnamefont {Boudouvis}}, \bibinfo
  {author} {\bibfnamefont {J.}~\bibnamefont {Lowengrub}}, \ and\ \bibinfo
  {author} {\bibfnamefont {I.~G.}\ \bibnamefont {Kevrekidis}},\ }\href@noop {}
  {\bibfield  {journal} {\bibinfo  {journal} {Phys. Rev. E},\ }\textbf
  {\bibinfo {volume} {85}},\ \bibinfo {pages} {031912} (\bibinfo {year}
  {2012})}\BibitemShut {NoStop}%
\bibitem [{\citenamefont {Wolfram}(1983)}]{Wolfram1983}%
  \BibitemOpen
  \bibfield  {author} {\bibinfo {author} {\bibfnamefont {S.}~\bibnamefont
  {Wolfram}},\ }\href@noop {} {\bibfield  {journal} {\bibinfo  {journal}
  {Rev.~Mod.~Phys.},\ }\textbf {\bibinfo {volume} {55}},\ \bibinfo {pages}
  {601} (\bibinfo {year} {1983})}\BibitemShut {NoStop}%
\bibitem [{\citenamefont {Azarmi}\ \emph {et~al.}(2002)\citenamefont {Azarmi},
  \citenamefont {Farid}, \citenamefont {Nokhodchi}, \citenamefont
  {Bahari-Saravi},\ and\ \citenamefont {Valizadeh}}]{Azarmi}%
  \BibitemOpen
  \bibfield  {author} {\bibinfo {author} {\bibfnamefont {S.}~\bibnamefont
  {Azarmi}}, \bibinfo {author} {\bibfnamefont {J.}~\bibnamefont {Farid}},
  \bibinfo {author} {\bibfnamefont {A.}~\bibnamefont {Nokhodchi}}, \bibinfo
  {author} {\bibfnamefont {S.~M.}\ \bibnamefont {Bahari-Saravi}}, \ and\
  \bibinfo {author} {\bibfnamefont {H.}~\bibnamefont {Valizadeh}},\ }\href@noop
  {} {\bibfield  {journal} {\bibinfo  {journal} {Int.~J.~Pharm.},\ }\textbf
  {\bibinfo {volume} {246}},\ \bibinfo {pages} {171} (\bibinfo {year}
  {2002})}\BibitemShut {NoStop}%
\bibitem [{\citenamefont {Chatterjee}\ \emph {et~al.}(2010)\citenamefont
  {Chatterjee}, \citenamefont {Rades},\ and\ \citenamefont
  {Tucker}}]{Chatterjee2}%
  \BibitemOpen
  \bibfield  {author} {\bibinfo {author} {\bibfnamefont {L.}~\bibnamefont
  {Chatterjee}}, \bibinfo {author} {\bibfnamefont {T.}~\bibnamefont {Rades}}, \
  and\ \bibinfo {author} {\bibfnamefont {I.~G.}\ \bibnamefont {Tucker}},\
  }\href@noop {} {\bibfield  {journal} {\bibinfo  {journal} {Int. J. Pharm.},\
  }\textbf {\bibinfo {volume} {384}},\ \bibinfo {pages} {87} (\bibinfo {year}
  {2010})}\BibitemShut {NoStop}%
\bibitem [{\citenamefont {Buchla}(2012)}]{codelocation}%
  \BibitemOpen
  \bibfield  {author} {\bibinfo {author} {\bibfnamefont {E.}~\bibnamefont
  {Buchla}},\ }\href@noop {} {\enquote {\bibinfo {title} {celldiff},}\
  }\bibinfo {howpublished} {available at
  \href{https://github.com/catfact/celldiff}{\texttt{https://github.com/catfact/celldiff}}}
  (\bibinfo {year} {August 2012})\BibitemShut {NoStop}%
\bibitem [{\citenamefont {Kuksal}\ \emph {et~al.}(2006)\citenamefont {Kuksal},
  \citenamefont {Tiwary}, \citenamefont {Jain},\ and\ \citenamefont
  {Jain}}]{Kuksal}%
  \BibitemOpen
  \bibfield  {author} {\bibinfo {author} {\bibfnamefont {A.}~\bibnamefont
  {Kuksal}}, \bibinfo {author} {\bibfnamefont {A.~K.}\ \bibnamefont {Tiwary}},
  \bibinfo {author} {\bibfnamefont {N.~K.}\ \bibnamefont {Jain}}, \ and\
  \bibinfo {author} {\bibfnamefont {S.}~\bibnamefont {Jain}},\ }\href@noop {}
  {\bibfield  {journal} {\bibinfo  {journal} {AAPS PharmSciTech},\ }\textbf
  {\bibinfo {volume} {7}},\ \bibinfo {pages} {E1} (\bibinfo {year}
  {2006})}\BibitemShut {NoStop}%
\bibitem [{\citenamefont {Wang}\ \emph {et~al.}(2002)\citenamefont {Wang},
  \citenamefont {Wang},\ and\ \citenamefont {Schwendeman}}]{Wang}%
  \BibitemOpen
  \bibfield  {author} {\bibinfo {author} {\bibfnamefont {J.}~\bibnamefont
  {Wang}}, \bibinfo {author} {\bibfnamefont {B.}~\bibnamefont {Wang}}, \ and\
  \bibinfo {author} {\bibfnamefont {S.}~\bibnamefont {Schwendeman}},\
  }\href@noop {} {\bibfield  {journal} {\bibinfo  {journal} {J.
  Control.~Release},\ }\textbf {\bibinfo {volume} {82}},\ \bibinfo {pages}
  {289} (\bibinfo {year} {2002})}\BibitemShut {NoStop}%
\bibitem [{\citenamefont {Wada}\ \emph {et~al.}(1995)\citenamefont {Wada},
  \citenamefont {Hyon},\ and\ \citenamefont {Ikada}}]{Wada}%
  \BibitemOpen
  \bibfield  {author} {\bibinfo {author} {\bibfnamefont {R.}~\bibnamefont
  {Wada}}, \bibinfo {author} {\bibfnamefont {S.}~\bibnamefont {Hyon}}, \ and\
  \bibinfo {author} {\bibfnamefont {Y.}~\bibnamefont {Ikada}},\ }\href@noop {}
  {\bibfield  {journal} {\bibinfo  {journal} {J. Control.~Release},\ }\textbf
  {\bibinfo {volume} {37}},\ \bibinfo {pages} {151} (\bibinfo {year}
  {1995})}\BibitemShut {NoStop}%
\end{thebibliography}%

\end{document}